\documentclass[aps,floats,prd,nofootinbib,twocolumn]{revtex4}

\usepackage{amssymb}
\usepackage{amsmath}

\usepackage{graphicx}
\usepackage{graphics}
\usepackage{dcolumn}
\usepackage{color}
\usepackage{fancyhdr} 
\usepackage{graphicx}

\usepackage{subfig}
\usepackage{bibshortcuts}

\usepackage[utf8]{inputenc}

\usepackage[justification=centerlast, format=plain, labelfont=bf]{caption}

\usepackage{hyperref} 
\hypersetup{
	colorlinks=true,
	linkcolor=blue,
	filecolor=magenta,      
	urlcolor=cyan,
}

\def\VEV#1{\left\langle #1 \right\rangle}

    \newcommand{\be}{\begin{equation}}
  \newcommand{\ee}{\end{equation}}
    \newcommand{\ba}{\begin{align}}
  \newcommand{\ea}{\end{align}}

\newcommand{\bse}{\begin{subequations}}
\newcommand{\ese}{\end{subequations}}

\begin{document}

\title{Efficient Computation of Galaxy Bias with Neutrinos and Other Relics}

\author{Julian B.~Mu\~noz\footnote{Electronic address: \tt julianmunoz@fas.harvard.edu}
} 
\author{Cora Dvorkin\footnote{Electronic address: \tt cdvorkin@g.harvard.edu }
}
\affiliation{Department of Physics, Harvard University, 17 Oxford St., Cambridge, MA 02138}

\date{\today}

\begin{abstract}
Cosmological data can be used to search for---and characterize---light particles in the standard model, if these populate our Universe. 
In addition to the well-known effect of these light relics in the background cosmology, usually parametrized through a change in the effective number $N_{\rm eff}$ of neutrino species, these particles can become nonrelativistic at later times, affecting the growth of matter fluctuations due to their thermal velocities.
An extensively studied example is that of massive neutrinos, which are known to produce a suppression in the matter power spectrum due to their free streaming.
Galaxies, as biased traces of matter fluctuations, can therefore provide us with a wealth of information about both known and unknown degrees of freedom in the standard model.
To harness this information, however, the galaxy bias has to be determined in the presence of massive relics, which is expected to vary with scale.
Here we present the code RelicFast, which efficiently computes the scale-dependent bias induced by relics of different masses, spins, and temperatures, through spherical collapse and the peak-background split.
Using this code, we find that, in general, the bias induced by light relics partially compensates the suppression of power, and should be accounted for in any search for relics with galaxy data.
In particular, for the case of neutrinos, we find that both the normal and inverted hierarchies present a percent-level step in the Lagrangian bias, with a size scaling linearly with the neutrino-mass sum, in agreement with recent simulations. 
This effect persists at the subpercent level even if one defines the Eulerian bias with respect to dark matter only, suggesting that it has to be properly included in cosmological searches for the neutrino mass.
RelicFast can compute halo bias in under a second, allowing for this effect to be properly included for  different cosmologies, and light relics, at little computational cost.
\end{abstract}

\maketitle

\section{Introduction}
\label{sec:Intro}

Cosmology can be a powerful tool in the search for physics beyond the standard model.
Specifically, new light particles, with weak couplings to the visible sector, are expected to decouple from it in the early Universe, while they are relativistic.
This freezes their distribution function, causing these particles to have a non-negligible thermal motions even at late times.
We will refer to these particles as light relics, and will parametrize them via their spin, mass, and temperature today.

Relics with very small masses contribute to the radiation energy density of our Universe at all times, which is commonly described as a change in the effective number $N_{\rm eff}$ of neutrino species present at a given era~\cite{Brust:2013xpv,Chacko:2015noa}.
Measurements of $N_{\rm eff}$ both during recombination~\cite{Ade:2015xua}, as well as during big-bang nucleosynthesis (BBN)~\cite{Cyburt:2015mya}, are in agreement with the $\Lambda$CDM prediction of $N_{\rm eff}=3.046$ within 10\%, thus constraining part of the light-relic parameter space.
Likewise, large-scale-structure (LSS) surveys are expected to reach similar sensitivities in $N_{\rm eff}$~\cite{Dodelson:2016wal,Baumann:2017gkg,Baumann:2018qnt}.

Massive relics can, in addition, leave striking cosmological signatures if they are nonrelativistic today, for which they are just required to have masses above an meV.
Let us take as an instance the case of neutrinos, which were relativistic when they decoupled, and are known to have a total mass of at least 60 meV~\cite{Fogli:2012ua,Abazajian:2013oma}, qualifying as massive relics. 
Their low mass guarantees that neutrinos have a significant thermal velocity throughout cosmic history, setting a free-streaming scale beyond which they do not cluster~\cite{Lesgourgues:2006nd}. 
As a consequence, small-scale fluctuations grow slower in a Universe with massive neutrinos, causing an observable suppression in the matter power spectrum.
Additionally, neutrinos change the background cosmology, producing a mismatch between high- and low-redshift measurements of the clustering of matter.
Current observations have constrained the sum of neutrino masses to be $\sum_i m_{\nu_i} \lesssim 0.2$ eV at 95\%~C.L., depending on the specific datasets considered and assumptions taken~\cite{Ade:2015xua,Vagnozzi:2017ovm,Abbott:2017wau},
and it is expected that the next generation of cosmological observables will provide a measurement of the sum of the neutrino masses~\cite{Abazajian:2013oma,Lesgourgues:2014zoa,Giusarma:2018jei,Pritchard:2008wy,Abazajian:2016yjj}.

Any massive relic will leave an imprint in matter fluctuations parallel to that of neutrinos.
As opposed to cosmologies with fully warm or fuzzy dark matter, in which the matter power spectrum nearly vanishes below some scale~\cite{Bode:2000gq,Hu:2000ke,Marsh:2015xka}, the presence of massive relics only causes a small suppression in the power spectrum, albeit at larger scales, which are easier to model.
Therefore, high-precision large-scale observables provide an ideal footing to search for these particles~\cite{Boyarsky:2008xj,Lesgourgues:2004ps,deBelsunce:2018xtd}.
Amongst these observables, measurements of clustering statistics of biased tracers, such as galaxies, are improving dramatically, and surveys like DESI~\cite{Levi:2013gra}, EUCLID~\cite{Amendola:2016saw}, and the LSST~\cite{Abate:2012za} will yield unprecedented measurements of galaxy power spectra.
In order to make progress, however, the galaxy bias has to be modeled (see Ref.~\cite{Desjacques:2016bnm} for a recent review), since massive relics are known to induce a scale dependence on this quantity, as pointed out in Ref.~\cite{LoVerde:2014pxa} for the case of neutrinos. 
Here we perform the first step towards that goal by calculating the linear galaxy bias in the presence of any massive relic.

We numerically solve the spherical collapse of haloes, taking into account the scale-dependent growth caused by the massive relics.
We have developed a software package, {\tt RelicFast}, which we make publicly available\footnote{ At: \url{https://github.com/JulianBMunoz/RelicFast}}.
Given the relic parameters, as well as the cosmological ones, {\tt RelicFast} provides the linear biases (Lagrangian and Eulerian) in a fraction of a second.
In particular, finding this bias is of critical importance in the search of neutrino masses, where the scale dependence in the bias partially compensates for the induced  suppression in power~\cite{LoVerde:2014pxa}, reducing it by a factor of $\sim 3$, depending on the neutrino and halo masses (we find, for instance, a reduction from $2\%$ to $0.5\%$ suppression for $\sum m_{\nu_i}=0.09$ eV for haloes of $M=10^{13}\, h \, M_\odot$).

Direct measurements of the scale dependence of the bias, for instance through cross correlations of CMB lensing and galaxy surveys, have been shown to be less sensitive to the neutrino mass than other observables~\cite{Schmittfull:2017ffw}. 
However, the different effects caused by neutrinos, and other relics, add or subtract coherently, so even if the scale dependence induced by neutrinos is not observable at high significance in isolation, it should be included when searching for these particles.
Additionally,  if any deviation on $N_{\rm eff}$ from the $\Lambda$CDM prediction was found in next-generation CMB studies, lower-redshift galaxy data would help to disentangle the characteristics of the particle sourcing it.

We note that galaxy bias is most reliably found through N-body simulations, or similar techniques~\cite{Scoccimarro:1997gr,Springel:2005mi,White:2013psd,Tassev:2013pn,Vogelsberger:2014dza,Feng:2016yqz,Hu:2016ssz}. 
However, there are several advantages to using a quasi-analytical approach, as the one we present here.
Firstly, it allows us to explore the parameter space more efficiently.
This might not be critical for the case of massive neutrinos, as there is one relevant parameter: the sum of neutrino masses~\cite{Liu:2017now,Villaescusa-Navarro:2013pva,Castorina:2013wga,Bird:2018all,AliHaimoud:2012vj,Chiang:2017vuk,Heitmann:2015xma,Bird:2011rb,Castorina:2015bma}; but in the case of light relics, both their masses and temperatures (or abundances) can vary, so any complete set of simulations would require a significant computational effort, whereas the bias for different relic cases can be found at low cost with {\tt RelicFast}.
Secondly, a great deal of intuition can be gained from quasi-analytic studies. For instance, we can easily find the galaxy bias for different cosmologies, allowing us to explore the degeneracies of the $\Lambda$CDM parameters with the light-relic degrees of freedom.
Thirdly, we are able to find the galaxy bias over a broader range of scales than commonly accessible to simulations, allowing us, for instance, to study wavenumbers both above and below the neutrino free-streaming scale.
Lastly, {\tt RelicFast} can run in a fraction of a second, which allows for a rapid change in input parameters (including cosmological ones), and  can therefore be implemented in any Markov-chain Monte Carlo search of light relics, including neutrinos.

In this paper we will show our formalism, and explore the capabilities of {\tt RelicFast}.
We start reviewing the spherical-collapse method in Section~\ref{sec:Collapse}, and comparing with the results from simulations for $\Lambda$CDM.
In Section~\ref{sec:Relics} we take a step back to describe the light relics, and their effects on linear perturbations, which we use in Section~\ref{sec:Relics_bias} to compute the halo bias and power spectrum for a universe with a light relic.
We, then, use the same methods
in Section~\ref{sec:Other_Relics} to explore the effects of general relics, including eV-mass sterile neutrinos and bosonic particles, and in Section~\ref{sec:Neutrinos} to study the scale-dependent bias caused by massive neutrinos.
Finally, we conclude in Section~\ref{sec:Conclusions}.

\section{Spherical Collapse and Bias}
\label{sec:Collapse}

We start by reviewing how to compute the bias of haloes in the spherical-collapse approximation, using the peak background-split argument~\cite{Fry:1992vr,Sheth:1999mn,Cole:1989vx}.
This section draws heavily from Ref.~\cite{LoVerde:2014pxa}, and readers familiar with the notation might want to skip to Section~\ref{sec:Relics}.
Throughout this work we set our fiducial cosmological parameters to a (physical) baryonic density $\Omega_b h^2=0.022$, dark-matter density $\Omega_d h^2=0.12$, $h=0.67$, and a nearly scale-free spectrum of primordial perturbations with an amplitude and tilt of $A_s=2.2\times 10^{-9}$ and $n_s=0.9655$, consistent with the values measured by the Planck collaboration~\cite{Ade:2015xua}, unless otherwise stated.
For convenience, we will also define the CDM+baryon (CDM+b) density as $\Omega_{c}   \equiv \Omega_d + \Omega_b$, 
and the Hubble parameter $H_0=100 \, h$ km s$^{-1}$ Mpc$^{-1}$.
We also assume that the three active neutrinos are massless, unless specified, contributing a total $N_{\rm eff}=3.046$ at the CMB epoch.
We set the dark-energy density $\Omega_{\Lambda}$ by requiring a flat Universe.

Schematically, we will obtain the bias of haloes of mass $M$ by using the peak background-split argument, and finding how a long-wavelength perturbation $\delta_L$ modulates their number density $n(M)$. 
We approach this problem by assuming that the halo, which was initially formed of a short-wavelength overdensity $\delta_S$ over a radius $R_{\rm ini}$, undergoes spherical collapse until it virializes. 
We then find the necessary $\delta_S$ to make it collapse at redshift $z_{\rm coll}$, as a function of the long-wavelength overdensity $\delta_L$.
However,  the halo mass function (HMF)---which tells us how many haloes of a certain mass there are---is generally a function of the critical overdensity $\delta_{\rm crit}$, which is obtained by extrapolating $\delta_S$ to the time of collapse. 
Thus, the collapse procedure provides us with $\delta_{\rm crit}$ as a function of the long-wavelength perturbation and, by assuming a functional form for the HMF, we will obtain the halo bias.

\subsection{Collapse}

Here, and throughout, we work in natural units, with $c=\hbar=k_B=1$.
We start with a halo of mass $M$ and radius $R(t)$. Assuming spherical collapse, its evolution is given by~\cite{LoVerde:2014pxa,LoVerde:2014rxa,Naoz:2005pd,Ichiki:2011ue,Schaefer:2007nf}
\be
\ddot{R}(t) = - \dfrac{G M}{R^2(t)} -\dfrac{4\pi G R(t)}{3} \sum_i [\rho_i(t)+3 P_i(t)] ,
\label{eq:Rdotdot}
\ee
where $G$ is Newton's constant, $\rho_i$ and $P_i$ are the energy density and pressure of species $i$, and the index $i$ runs over all non-CDM+baryon species.
Once a starting redshift is selected, which we choose at $z_{\rm ini}=200$, we can compute the average (physical) size of haloes of mass $M$ at that redshift as
\be
\bar{R}_{\rm ini} = \left(\dfrac{H_0^2 \Omega_{c}}{2\,G M}\right)^{-1/3}  (1+z_{\rm ini})^{-1}.
\ee
We obtain the initial conditions for Eq.~\eqref{eq:Rdotdot} in the presence of both long- and short-wavelength CDM+b perturbations ($\delta_L$ and $\delta_S$, respectively),  as
\be
R_{\rm ini} = \bar{R}_{\rm ini}\left(1 -  \dfrac{\delta_S + \delta_L}{3} \right),
\ee
and 
\be
\dot R_{\rm ini} = {R}_{\rm ini} \left(H(z_{\rm ini}) -  \dfrac{\dot \delta_S + \dot \delta_L}{3} \right),
\ee
where the Hubble parameter is given by
\be
H(z) = \dfrac{8\pi G}{3} \sqrt{\rho_{c} (z) + \sum_i \bar \rho_i(z)},
\label{eq:Hubble}
\ee
$\rho_{c}(z)$ is the CDM+b density at redshift $z$, and $\bar \rho_i(z)$ is the spatial average of $\rho_i(z)$. We emphasize that $i$ includes all non-CDM+b species, and $\delta_L$ and $\delta_S$ are always evaluated at $z_{\rm ini}$.

There are two simple ways to obtain the small-scale perturbation velocity, $\dot \delta_S$, from  $\delta_S$.
The first is through the variance of fluctuations in the scale of the halo,
\be
\sigma^2(M,z) = \int \dfrac{dk\,k^2}{2\pi^2} P_{cc} (k) W^2(k R_M),
\label{eq:sigma_M}
\ee
where $P_{cc}(k)$ is the CDM+b power spectrum, obtained from the baryon ($b$) and CDM $(d)$ power spectra through
\be
\Omega_c P_{cc} = \Omega_d P_{dd} + \Omega_b P_{bb},
\ee
and we choose a top-hat window function $W(x) = 3[\sin(x)/x-\cos(x)]/x^2,$
defining the comoving halo radius as
\be
R_M \equiv \left(\dfrac{H_0^2 \Omega_{c}}{2 \,G M}\right)^{-1/3}.
\ee
In this case we can set the initial perturbation velocity as
\be
\dfrac{\dot \delta_S}{\delta_S} = \dfrac{\dot \sigma(M,z_{\rm ini})}{\sigma(M,z_{\rm ini})}.
\ee
We could, instead, set the velocity through
\be
\dfrac{\dot \delta_S}{\delta_S} = \dfrac{\dot{\mathcal T}_c(k_*,z_{\rm ini})}{\mathcal T_c(k_*,z_{\rm ini})},
\label{eq:transfer_IC}
\ee
where $\mathcal T_c(k,z)$ is the CDM+b transfer function, obtained from a Boltzman code, like {\tt CLASS}~\cite{Blas:2011rf} or {\tt CAMB}~\cite{Lewis:1999bs},
and $k_*= \pi/R_M$ is chosen to match the scale of the halo. We find that these two methods produce nearly identical results, and we will use the first one throughout this work.

In the presence of a long-wavelength CDM+b perturbation $\delta_L (k)$, with a wavenumber $k$, the rest of components see their densities and pressures modulated as
\begin{subequations}
	\label{eq:deltarho}
	\ba
	\rho_i(z) =&\, \bar{\rho}_i(z) \left[1 + \delta_i(z)\right], \\
	P_i(z) =&\, \bar{P}_i(z) \left[1 + \dfrac{c_{s,i}^2(z)}{w_i(z)} \delta_i(z)\right],
\end{align}
\end{subequations}
where $\bar P_i (z) = w_i(z) \bar{\rho}_i(z)$ is the spatially averaged pressure of component $i$, $w_i(z)$ is its equation of state, and $c_{s,i}$ is its sound speed. 
For convenience we have defined 
\be
\delta_i(z) \equiv \delta_L  \dfrac{\mathcal T_i(k,z)}{\mathcal T_c(k,z_{\rm ini})},
\ee
\newline
where $\mathcal T_i$ is the transfer function of the $i$-th component. 
Note that, as opposed to Ref.~\cite{LoVerde:2014pxa}, we properly incorporate the nonvanishing sound speed $c_{s,i}$ for all species we study, including massive neutrinos. 
In the neutrino case, on which we will focus on Section~\ref{sec:Neutrinos}, we find that artificially setting $c_{s,i}^2=0$ (and thus ignoring pressure fluctuations) overestimates the scale-dependence of the bias by as much as a factor of two.
This is perhaps not surprising, as for free-streaming relics $c_{s,i}^2\sim w_i$, so pressure and density fluctuations can be comparable on the right-hand side of Eq.~\eqref{eq:Rdotdot}.
We note that this issue does not arise when evolving $\delta_S(z)$ instead of $R(z)$, as done for instance in Ref.~\cite{Chiang:2017vuk}, where the information on the sound speed of all components is contained in the evolution of the long-wavelength CDM+b mode.

The procedure consists of solving for $R(z)$, given a fixed cosmology and halo mass $M$, and varying $\delta_S$ until the halo collapses ($R\to0$) at our chosen redshift $z_{\rm coll}$. 
In reality, of course, haloes virialize and possess a finite radius at collapse, although solving for $R\to0$ is a good proxy for virialization~\cite{LoVerde:2014rxa,LoVerde:2014pxa}.
For computational simplicity we solve Eq.~\eqref{eq:Rdotdot} using redshift, as opposed to physical time, as a coordinate. We detail in Appendix~\ref{app:Code} the coordinate transformation required.
We will repeat this procedure with different values of $\delta_L$, in order to find $\delta_S(\delta_L)$. We obtain the critical overdensity as
\be
\delta_{\rm crit} = \delta_S  \dfrac{ \sigma(M,z_{\rm coll})}{\sigma(M,z_{\rm ini})},
\ee
although if we had chosen to set the initial conditions with Eq.~\eqref{eq:transfer_IC} instead, we would find the critical overdensity with $\mathcal T(k_*,z_{\rm coll})/\mathcal T(k_*,z_{\rm ini})$.
Finally, we also evolve $\delta_L$ to the redshift of collapse to find
\be
\delta_{L,\rm coll} (k) =  \delta_L (k)  \dfrac{\mathcal T_c(k,z_{\rm coll})}{\mathcal T_c(k,z_{\rm ini})}.
\ee

\subsection{Bias}

Given the resulting function $\delta_{\rm crit}[\delta_{L,\rm coll}(k)]$, we can find the linear Lagrangian bias, with respect to CDM+b, through the peak background-split argument~\cite{Fry:1992vr,Sheth:1999mn}, to be
\be
b_1^L(k)  = \left . \left(\dfrac{\partial \log n}{\partial \delta_{\rm crit}}\right) \right |_{\delta_{L,\rm coll}=0} \left(\dfrac{d \delta_{\rm crit}}{d \delta_{L,\rm coll}(k)}\right),
\label{eq:b1Ldef}
\ee
where we assume that the only change to the halo mass function (HMF) is through $\delta_{\rm crit}$, and its functional form is otherwise unaltered by any new particles.
To perform this calculation we need to assume a shape of the HMF.
We use a fit to the MICE simulations of Ref.~\cite{Crocce:2009mg}, which has been shown to yield a good approximation to the mass function even in the presence of light relics (massive neutrinos)~\cite{Castorina:2013wga}. 
The derivative for the HMF that we take is then
\be
\dfrac{\partial \log n}{\partial \delta_{\rm crit}} = -\dfrac{2 c(z) \delta_{\rm crit}}{\delta_{\rm ref}^2  \sigma^2} + \dfrac{a(z)}{\delta_{\rm crit}\,[1+b(z)\,(\delta_{\rm ref}  \sigma/\delta_{\rm crit})^{a(z)}]},
\label{eq:HMF}
\ee
obtained by performing the transformation $\sigma\to\sigma \, \delta_{\rm ref}/\delta_{\rm crit}$ to the fit in Ref.~\cite{Crocce:2009mg},  with $\delta_{\rm ref}=1.686$, and with parameters
$a(z)=1.37\,(1+z)^{-0.15}$, $b(z)=0.3\,(1+z)^{-0.084}$, and $c(z)=1.036\,(1+z)^{-0.024}$.
We obtain the CDM+b variance $\sigma^2$ with Eq.~\eqref{eq:sigma_M}.

For completeness, we have also implemented the result for the HMF from Refs.~\cite{Sheth:1999mn} and \cite{Bhattacharya:2010wy}. 
The scale dependence of $b_1^L$ is, by construction, independent of the chosen HMF, as the first term in Eq.~\eqref{eq:b1Ldef} is evaluated at $\delta_{L,\rm coll}=0$ and thus does not depend on $k$.
The three HMFs produce, however, different normalizations of $b_1^L$ at the percent level, which has a negligible impact on all the scale dependences that we study in this work, as we show in Appendix~\ref{app:HMF}.

We now move on to compute the Eulerian bias.
By transforming from Lagrangian space (defined by the CDM+b fluid) to Eulerian space, we find the halo overdensity
\be
\delta_h = (1+b^L_1) \delta_c,
\ee
in terms of the CDM+b overdensity $\delta_c$.
This is to be compared with the equivalent definition as a function of the matter fluctuation $\delta_m$,
\be
\delta_h = b_1 \delta_m,
\ee
where $b_1$ is the linear Eulerian bias, which we can then easily find to be
\be
b_1 =\dfrac{P_{hm}}{P_{mm}} = \sqrt{\dfrac{P_{hh}}{P_{mm}}},
\label{eq:b1}
\ee
where $P_{hh}$ is the halo (auto) power spectrum, $P_{mm}$ is the matter power spectrum, and $P_{hm}$ is the halo-matter cross spectrum.
Throughout this work  we will often drop the ``Eulerian" label and refer to $b_1$ simply as linear bias, unless confusion can arise.

When adding relics, we will account for a number $N_{\rm sp}$ of matter species, including CDM+b. Then,
we calculate the matter power spectrum as
\be
P_{mm} = \sum_{i,j}^{N_{\rm sp}}  f_i f_j P_{ij},
\ee
where 
$f_i=\Omega_i/\Omega_m$ are the fractions of the total matter in each component today, given that $\Omega_i$ is their abundance and $\Omega_m \equiv \sum_i \Omega_i$, and $P_{ij}$ are their power/cross spectra.
Additionally, the halo-matter cross spectrum is found as
\be
P_{hm} = (1+b^L_1) \sum_i^{N_{\rm sp}}   f_i P_{ci}.
\ee
As an example, in the case of a cosmology with CDM+b and a light relic, which carries a fraction $f_X$ of the total matter, the matter power spectrum is given by
\be
P_{mm} = (1- f_X)^2 P_{cc} + 2  f_X (1- f_X) P_{cX} +  f_X^2P_{XX},
\ee
whereas the halo-matter cross spectrum is 
\be
P_{hm} = (1+b^L_1) \left[(1- f_X) P_{cc} +  f_X P_{cX}\right].
\ee
The ratio of these two quantities gives us the linear Eulerian bias.

Before moving on to specific realizations of halo bias, let us end this discussion with two cautionary remarks. 
First, the spherical-collapse model we employ is expected to be a good approximation for massive haloes, as smaller haloes are expected to deviate from sphericity~\cite{Eisenstein:1994ni,Sheth:1999su,Sheth:2001dp}.
Additionally, different effects, such as those from a tidal shear, can cause a dispersion in the barrier for collapse ($\delta_{\rm crit}$)~\cite{Baldauf:2012hs,Castorina:2016tuc}, which we ignore here.
This, nonetheless, is expected to affect less the bias of smaller-mass haloes~\cite{Paranjape:2017zpc}.
Second, we are using haloes as proxies for galaxies~\cite{Wechsler:2018pic}, and we are not including any information on their environment, or assembly history, which are known to produce additional biases and stochasticity\footnote{Stochasticity causes a difference between $P_{hh}$ and $P_{hm}^2/P_{mm}$, which our formalism neglects.}~\cite{Wechsler:2001cs,Dalal:2008zd,Chue:2018hxk,Baldauf:2013hka}.
Nevertheless, our calculation suffices to show the scale-dependent effect of light relics in the galaxy power spectrum, and we do not expect any of the aforementioned effects to substantially change this behavior in the linear regime.
This has been confirmed for the case of massive neutrinos, where a spherical-collapse calculation has been shown to agree with N-body simulations~\cite{Chiang:2017vuk}.
In addition, massive neutrinos have also been shown to also induce a scale-dependent bias in voids~\cite{Banerjee:2016zaa}.
We leave refining the calculation by studying the effects of environment and non-spherical collapse in the calculation for future work.

\subsection{An Example}
\label{sec:Example}

We will illustrate the procedure in the simplest scenario of $\Lambda$CDM with massless neutrinos. 
In this case there are three components that contribute to the right-hand side of Eq.~\eqref{eq:Rdotdot}: photons (with $w_\gamma = c_{s,\gamma}^2=1/3$ and temperature $T_\gamma^{(0)}=2.73$ K today), massless neutrinos (also with $w_\nu=c_{s,\nu}^2=1/3$, but with $T_\nu^{(0)} = 1.95$ K), and dark energy (with $w_\Lambda=-1$, no fluctuations, and energy density given by the closure equation).

\begin{figure}[hbtp!]
	\includegraphics[width=0.44\textwidth]{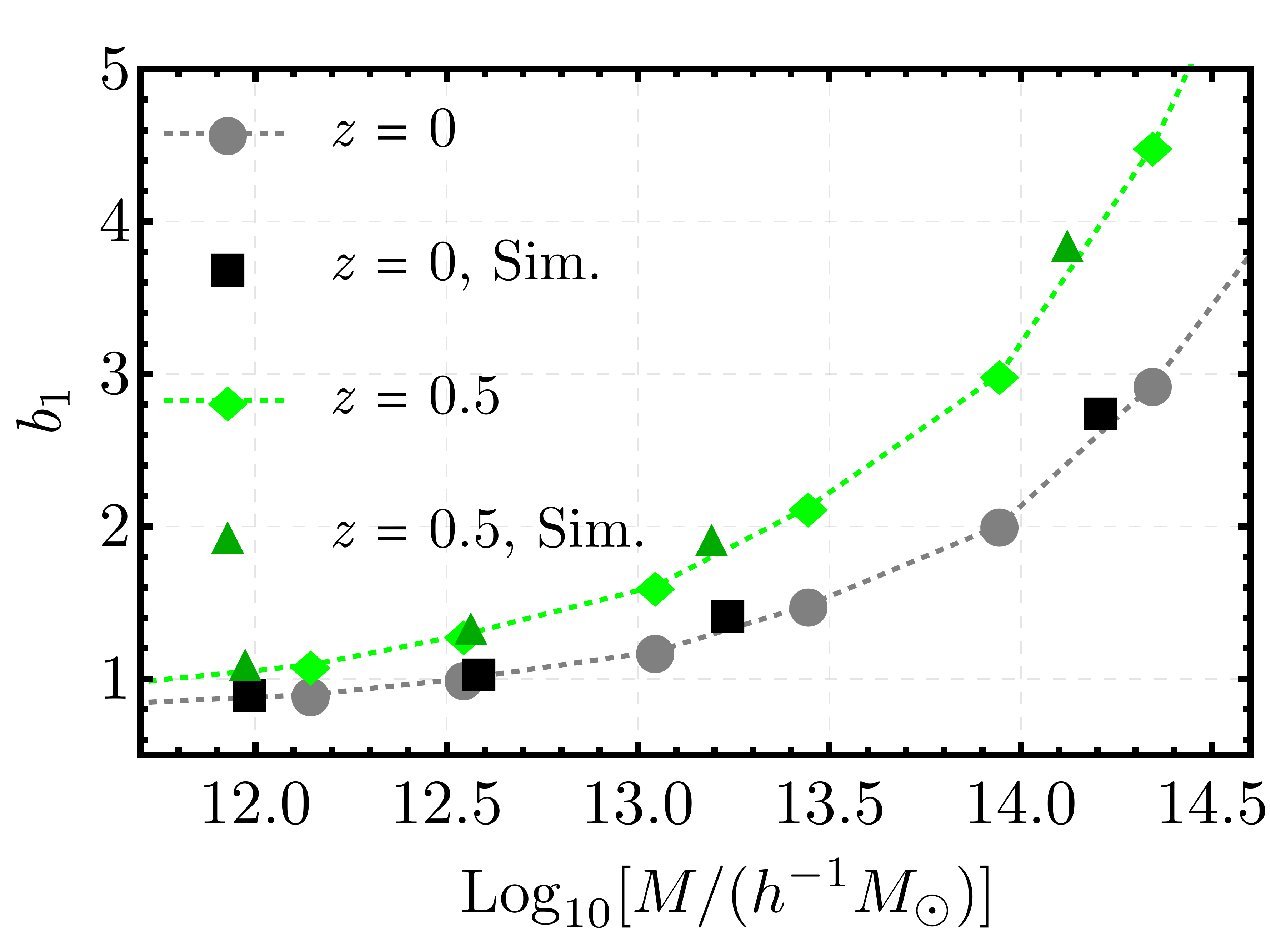}			
	\caption{Linear (Eulerian) bias of haloes of different masses $M$, at two redshifts, computed with {\tt RelicFast}, where we have joined the points for visual aid. 
	We have adopted a cosmology consistent with the MICE simulations, as explained in Section~\ref{sec:Example}.
	The  results of simulations from Ref.~\cite{Hoffmann:2016omy} are shown as black squares for $z=0$ and dark-green triangles for $z=0.5$.
	}
	\label{fig:b_M_LCDM}
\end{figure}

We find the Eulerian bias trivially from the Lagrangian one through $b_1 = (1  + b_1^L)$, as CDM+b is the only matter fluid (since massless neutrinos are radiation even at $z=0$), and we show it in Fig.~\ref{fig:b_M_LCDM} for two different redshifts, as a function of the mass $M$ of the halo.
Here we have set the long-wavelength mode to be $k=10^{-3}$ Mpc$^{-1}$, although the results do not depend sensitively on this number. This Figure shows the well-known results that, at a given redshift, heavier haloes are more biased, as they are harder to form;
and that, for any given halo mass, the bias increases with redshift, as one needs larger overdensities to collapse earlier.
Moreover, we compare our results with the bias measurements from two-point halo-matter
cross-correlations on simulations from Ref.~\cite{Hoffmann:2016omy}, at both redshifts, finding excellent agreement.
For this Figure we have modified our fiducial cosmology to $h=0.7$, $n_s=0.966$, and $\Omega_d h^2=0.10$, in order to match that of Ref.~\cite{Hoffmann:2016omy} and the MICE simulations\footnote{ 	\url{ http://maia.ice.cat/mice/}}~\cite{Crocce:2009mg}.

\section{Light Relics}
\label{sec:Relics}

Now that we have outlined the procedure to obtain the halo bias, let us describe the properties of light relics, on which will focus for the rest of this work.
Our motivation to study this case is twofold.
First, many extensions of the standard model predict light degrees of freedom, which could be in thermal contact with the visible sector in the early Universe. These would leave a cosmological imprint as they contribute to the cosmic energy budget.
Current and upcoming galaxy surveys can be sensitive to intermediate-mass relics, which might otherwise be inaccesible with CMB data.
Second, neutrino-oscillation experiments have shown hints for an eV-mass sterile neutrino~\cite{Gariazzo:2015rra,Abazajian:2017tcc}, which could compose part of the dark matter~\cite{Hamann:2011ge,Dasgupta:2013zpn}.
Galaxy power spectra can, therefore, settle the issue of whether sterile neutrinos are cosmologically present.

\subsection{Cosmology of Light Relics}

We begin with a brief review of the cosmology of light relics.
Particles that decouple while being relativistic keep their distribution function intact, with their temperature $T_X$ given by that of photons, $T_\gamma$, at the time of decoupling, and redshifting simply as $T_X(z) = (1+z)\,T_X^{(0)}$, where $T_X^{(0)}$ is their temperature today. 
Light relics can, as opposed to adiabatically cooling degrees of freedom, exhibit large thermal motion, even at low redshifts.
The temperature of relics today is not necessarily the same as that of photons, since photons are heated by the annihilation of standard-model degrees of freedom.
For instance, neutrinos started decoupling before electron-positron annihilation, when $T_\gamma\sim$ MeV, which causes the neutrino temperature today to be a roughly a factor of $(4/11)^{1/3}$ smaller than that of photons.
Previous to that, no significant heating is expected to occur until $T_\gamma\sim 200$ MeV, when the QCD phase transition erased a myriad of degrees of freedom,
so any relic that decoupled between BBN and the QCD phase transition would roughly have the same temperature today as neutrinos, $T_X^{(0)}\sim$ 2 K.
Relics that decoupled before (or during) the QCD phase transition would be colder, reaching temperatures today as low as $T_X^{(0)}\sim$ 1 K~\cite{Brust:2013xpv}.
Therefore, the range $T_X^{(0)}=[1-2]$ K brackets the reasonable values of relic temperatures, unless large amounts of new (and unknown) degrees of freedom are active in the very early Universe.

We will be agnostic about the origin of the relics, and parametrize any new degree of freedom $X$ through its mass $m_X$ and temperature $T_X^{(0)}$ at redshift zero.
We assume that these particles are part of one family with two spin-1/2 degrees of freedom (just like active neutrinos), which decoupled while relativistic, so they keep a Fermi-Dirac distribution
\be
f^{\rm FD}_X(q,z) = \dfrac{1}{e^{\,q/T_X(z)}+1},
\ee
where $q$ is their momentum and $T_X(z) = T_X^{(0)}\,(1+z)$.
Then, their energy density and pressure are given by
\begin{subequations}
\ba
\bar \rho_X(z) &= 2 \int \dfrac{d^3q }{(2\pi)^3}\, E(q) f^{\rm FD}_X(q,z),  \\
\bar P_X(z) &= 2 \int \dfrac{d^3q }{(2\pi)^3} \dfrac{\, q^2 f^{\rm FD}_X(q,z)}{3 E(q) },
\end{align}
\end{subequations}
where we have defined $E(q) \equiv \sqrt{q^2 + m_X^2}$ for convenience. 
From these two parameters we can find their equation of state as
\be
w_X(z) = \dfrac{\bar P_X(z)}{\bar \rho_X(z)}
\label{eq:EoS_X}
\ee
and their abundance as $\Omega_X = \bar \rho_X(z=0) /\rho_{\rm crit}$, where $\rho_{\rm crit}$ is the critical energy density. For this one-family case, the light-relic abundance can be well approximated by\footnote{In {\tt RelicFast} we use $T_\nu^{(0)}/ T_\gamma^{(0)} = 0.71599$, known to be a better approximation to the non-instantaneous neutrino decoupling than $(4/11)^{1/3}$~\cite{Mangano:2005cc,Lesgourgues:2011rh}. This dictates the value of the denominator in Eq.~\eqref{eq:omega_X}.}
\be
\Omega_X h^2 \approx \dfrac{m_X}{93.14 \, \rm eV} \left(\dfrac{T_X^{(0)}}{T_\nu^{(0)}}\right)^3.
\label{eq:omega_X}
\ee
We will relax these assumptions later, and show how other light relics, even with integer spins, can be expressed in terms of an ``equivalent neutrino" given $m_X$ and $\Omega_X$.

Additionally, the fluid-like nature (or lack thereof) of a light relic  determines its sound speed and viscosity.
It has been argued that it can be distinguished whether light relics behave as a fluid or stream freely, through their effect on the phase of the acoustic peaks~\cite{Bashinsky:2003tk,Follin:2015hya,Baumann:2015rya,Baumann:2017lmt,Choi:2018gho}.
We will assume that all light relics, including neutrinos, have no important interactions, and thus are freely streaming.
In this case we can write the  sound speed of a relic as
\be
c_{s,X}^2 \approx c_{{\rm ad},X}^2 = \dfrac{\dot{\bar P}_X}{\dot{\bar \rho}_X},
\label{eq:cad}
\ee
where $c_{{\rm ad},X}$ is their adiabatic sound speed, which is a good approximation to the sound speed at a lower computational cost~\cite{Lesgourgues:2011rh}.
Likewise, we will assume that relics have the usual viscosity due to their freely streaming nature, which does not enter our formalism, although it can be modified in {\tt CLASS}.

\subsection{Current Constraints}

Any new relavistic particles alter the rate of expansion, as they behave as radiation, which can be constrained with CMB anisotropies and with measurements of the cosmic abundances resulting from BBN.
We can write the radiation energy density as
\be
\rho_R = \dfrac{\pi^2}{15} T_\gamma^4\left[1+\dfrac{7}{8} \left(\dfrac{T_\nu}{T_\gamma}\right)^{4/3} N_{\rm eff}\right],
\ee
where the $7/8$ factor arises because of the fermionic nature of neutrinos. 
We can parametrize new light degrees of freedom through their contribution to $N_{\rm eff}$ at both the CMB and BBN epochs. 
The non-instantaneous decoupling of neutrinos leaves a signature in the effective number $N_{\rm eff}$ of neutrino species, which has the value $N_{\rm eff}=3.046$ in the standard model~\cite{Mangano:2005cc}.
Current Planck data of CMB temperature anisotropies, plus large-scale polarization information (TT+lowP), can constrain deviations from this prediction to be $|\Delta N_{\rm eff}^{\rm CMB}|<0.3$ within 68\% C.L.~\cite{Ade:2015xua}, and
the upcoming CMB-S4 experiment is expected to improve this figure by an order of magnitude~\cite{Abazajian:2016yjj}.
However, relics with masses above an eV will not fully contribute to $\rho_R$ at decoupling, rendering them difficult to constrain with CMB measurements alone. 
All relics that we study are, nonetheless, relativistic during BBN (if they were present in that era), where the 1-$\sigma$ constraint is $|\Delta N_{\rm eff}^{\rm BBN}|<0.3$~\cite{Cyburt:2015mya}.

As long as new particles are relativistic, their contribution to  $N_{\rm eff}$ depends solely on their temperature and degrees of freedom, $g_X$, and can be written as
\be
\Delta N_{\rm eff} = \dfrac{2\,g_X}{3} \left( \dfrac{T_X^{(0)}}{T_\nu^{(0)}} \right)^4.
\ee
Requiring $\Delta N_{\rm eff}<0.3$ at BBN thus forces $T_X^{(0)} \leq  1.4$ K for a new neutrino-like family, with $g_X=3/2$.
We are interested in relics that are nonrelativistic today, i.e., $m_X \gg T_X^{(0)}$, which requires $m_X \gtrsim$ meV.
For ease of visualization we will express our results in terms of the relic mass $m_X$ and fraction $f_X\equiv \Omega_X/\Omega_m$, where $\Omega_m \equiv \Omega_c + \Omega_X$ is the total matter density (and we remind the reader that $\Omega_c$ is the CDM+b density).

\subsection{Effect on the Matter Power Spectrum}

The effects of light relics on matter perturbations can be divided in two broad categories, those caused by a mismatch in the definition of (clustering) matter at small scales, and those caused by feedback on the rest of the matter. 

The relics we are studying are included as matter in the cosmic inventory, so the total matter perturbations at low redshift are sourced by both CDM+b ($c$) and light relics $(X)$, with
\be
\delta_m =  f_c \delta_c +  f_X \delta_X,
\ee
where $f_c =\Omega_c/\Omega_m$.
As opposed to CDM, light relics can have enough thermal velocities to stream out of potential wells.
In analogy with the case of neutrinos, we can approximate the small-scale light-relic perturbations as~\cite{Lesgourgues:2006nd}
\be
\delta_X (k\gg k_{\rm fs}) \sim \left( \dfrac{k}{k_{\rm fs}}\right)^{-2} \delta_m,
\ee 
where $k_{\rm fs}$ is the relic free-streaming scale, given by~\cite{AliHaimoud:2012vj}
\ba
k_{\rm fs} (z) &=  \left(\dfrac{3}{2} \VEV{v_X^{-2}(z)}_{\rm FD} \right)^{1/2}\dfrac{H(z)}{(1+z)} \nonumber\\
&\approx \dfrac{0.08}{\sqrt{1+z}} \left(\dfrac{m_X}{0.1\,\rm eV}\right) \left(\dfrac{T_X^{(0)}}{T_\nu^{(0)}}\right)^{-1} \,h \,\rm Mpc^{-1},
\label{eq:q:kfs}
\end{align}
assuming matter domination and our fiducial cosmology.
Here we have assumed that $X$ are nonrelativistic particles with temperature $T_X$, following a Fermi-Dirac distribution, so  $\VEV{v^{-2}(z)}_{\rm FD}= 2\log(2)m_X^2 [3\zeta(3)\,T_X^2(z)]^{-1}$.
Therefore, just from the absence of $X$ perturbations for $k>k_{\rm fs}(z)$, the matter power spectrum will be suppressed by a factor of $(1-f_X)^2$.

Additionally, since relics do not cluster in small scales, the growth of CDM perturbations is stunted. This back reaction further suppresses the small-scale matter fluctuations.
It is estimated that in the presence of massive neutrinos (carrying a fraction $f_\nu$ of the total matter) the CDM overdensities evolve as $\delta_c \propto a^{1-3 f_\nu/5}$, as opposed to $\delta_c\propto a$~\cite{Lesgourgues:2006nd}.
This lower growth rate, combined with the missing neutrino fluctuations at small scales, yields the well-known (linear) result that $\Delta P_{mm}/P_{mm} \approx 1-8  f_\nu$ for $k\gg k_{\rm fs}$ (similarly, $\Delta P_{cc}/P_{cc} \approx 1-6  f_\nu$).

For non-neutrino relics these results are slightly different.
If a light relic becomes nonrelativistic during radiation domination, as most of the cases we study do, the suppression of power starts at the free-streaming horizon, defined as~\cite{Boyarsky:2008xj}
\be
k_{\rm fsh} (t_0) = \left[ \int_0^{t_0} dt \VEV{v}/a(t) \right]^{-1} \sim \sqrt{\Omega_R} H_0/\VEV{v(t_0)},
\ee
where $\Omega_R$ is the energy density in radiation today, and $ \VEV{v}$ is the averaged velocity of the relics, roughly given by $c$ before the particles turn nonrelativistic, and by $3\,T_X(z)/m_X$ afterwards.
This wavenumber can be significantly smaller than $k_{\rm fs}$, since the free-streaming horizon $k_{\rm fsh}$ keeps shrinking during radiation domination (after the relics become nonrelativistic), whereas $k_{\rm fs}$ does not.
Similarly, the suppression in the CDM growth factor in the presence of light relics is more pronounced than for neutrinos, if these relics become nonrelativistic during radiation domination, and can be approximated by $\delta_c \propto a^{1-3 f_X/4}$~\cite{Boyarsky:2008xj} (cf. the neutrino exponent of $1-3 f_\nu/5$). Therefore, for small values of $f_X$, the suppression for $k \gg k_{\rm fs}$ is given by $\Delta P_{mm}/P_{mm} = (1-14  f_X)$, and is thus much larger than for neutrinos carrying the same fraction of matter.
We will obtain the transfer functions of these particles from the publicly available {\tt CLASS} code~\cite{Blas:2011rf,Lesgourgues:2011rh}, which numerically includes all these effects at high precision.

For completeness, we will also define the nonrelativistic scale as the wavenumber that crossed the horizon when the relics became nonrelativistic, i.e., 
\be
k_{\rm nr} = a_{\rm nr} H(a_{\rm nr}),
\ee
with $a_{\rm nr}=T_X^{(0)}/m_X$, roughly corresponding to the scale factor at which half of the Fermi-Dirac distribution would have $p<T_X$~\cite{AliHaimoud:2012vj}.
Given that all these scales are a function of $T_X^{(0)}/m_X$, and only have mild redshift dependences, we choose to parametrize our results in terms of $k_{\rm fs}$ for simplicity.

\section{Bias from Light Relics}
\label{sec:Relics_bias}

So far we have only discussed the effect of light relics on the matter power spectrum. 
Let us now move on to calculate their effect on the galaxy bias.

The free streaming of light relics stunts the development of linear perturbations, which manifests itself as a scale-dependent growth.
This is known to cause a scale-dependent bias in general~\cite{Hui:2007zh,Parfrey:2010uy}, and in particular for neutrinos~\cite{LoVerde:2014pxa,LoVerde:2016ahu}.
Intuitively, the scale dependence of the bias arises from the sensitivity of halo formation to the history of perturbation growth~\cite{Senatore:2014eva,Chiang:2017vuk}.
To see why, let us compare two very different scenarios, remembering that the bias is defined as the (logarithmic) change in the abundance of haloes in the presence of a long-wavelength perturbation.
In the first scenario, the growth of perturbations is simply given by a scale-independent growth factor $D_+(z)$, and thus, chosen some long-wavelength perturbation at collapse $\delta_{L,\rm coll}$, its value at previous redshifts is $\delta(z) = \delta_{L,\rm coll} D_+(z)/D_+(z_{\rm coll}) $.
In the second scenario, the perturbation is frozen at some small value until some redshift $z_*$, after which it quickly transitions to its value today, so $\delta(z) \approx \Theta(z_*-z) \delta_{L,\rm coll}$.
In the first scenario increasing $\delta_{L,\rm coll}$  produces additional fluctuations at all previous times, which significantly impacts the abundance of haloes. In the second scenario, however, for small-enough values of $z_*$, changing $\delta_{L,\rm coll}$ has little effect on the halo abundance. Thus, even though
these two cases share the same $\delta_{L,\rm coll}$, the change in the number of haloes in the presence of a long-wavelength perturbation depends on the growth history of this perturbation, which causes different halo biases.
Operationally, the bias becomes scale dependent due to the change in $\delta_{\rm crit}$ with different long-wavelength perturbations $\delta_{L,\rm coll} (k)$, as seen in Eq.~\eqref{eq:b1Ldef}.
This derivative depends both on the CDM+b transfer functions at $z_i$ and $z_{\rm coll}$, which are used to find $\delta_{\rm crit}$ and $\delta_{L,\rm coll}$ from the initial $\delta_S$ and $\delta_L$, as well as on the transfer functions of the other components at all intermediate redshifts, through the long-wavelength perturbation of non-CDM fluids.
In fact, even in $\Lambda$CDM (with massless neutrinos) there is a small growth difference between modes that entered the horizon before and after matter-radiation equality. This difference, added to the effect of photon and massless-neutrino perturbations during the halo collapse, yields a small scale-dependence of the bias, which will become apparent in our analysis.

We emphasize that even if the scale dependence of the bias induced by light relics is not observable at high significance in isolation~\cite{Schmittfull:2017ffw,OurForecasts}, 
it partially counteracts the suppression that these particles produce, 
so it is imperative to characterize it.

\subsection{Light-Relic Clustering}

The thermal velocity of light relics is finite, so some of them can accumulate in DM haloes.
This has been extensively studied for the case of neutrinos,
which form ``fuzzy" neutrino haloes, more loosely bound than the DM haloes, and thus more extended~\cite{Brandenberger:1987kf,Ringwald:2004np,AliHaimoud:2012vj,LoVerde:2013lta}.
This effect is most important for cluster-sized haloes, with $M\sim 10^{15}\, M_\odot$, which have deeper potential wells, and for heavier neutrinos, with masses $m_\nu\gtrsim 1$ eV, and thus lower velocities.

The properties of a putative neutrino halo around the Milky Way can affect, for instance, direct-detection efforts of the cosmic neutrino background~\cite{deSalas:2017wtt}. 
Here, however, we are only interested in the overall effect of relics on the spherical collapse of haloes. 
Relic clustering can be accounted for through a new term in Eq.~\eqref{eq:Rdotdot}, which now reads~\cite{LoVerde:2014rxa,Ichiki:2011ue}
\be
\ddot{R}(t) = - \dfrac{G [M + \delta M_X (t)]}{R^2(t)} -\dfrac{4\pi G R(t)}{3} \sum_i [\rho_i(t)+3P_i(t)] ,
\label{eq:Rdotdot_Clust}
\ee
where $\delta M_X(t)$ is the amount of accreted light-relic mass within $R(t)$.
We detail our procedure to obtain $\delta M_X(t)$ in Appendix~\ref{app:Clustering}, using the first-order ``BKT" approximation from Ref.~\cite{Brandenberger:1987kf}.
In Ref.~\cite{LoVerde:2014rxa} it was explored what is the change in the halo collapse when including neutrino clustering, and it was found that using this BKT approximation to find $\delta M_X(t)$ reproduced the $\delta_{\rm crit}$ from an N-1-body simulation with good accuracy, even for cluster-sized haloes.
Throughout this work we will focus on haloes with $M \sim 10^{13} M_\odot$, where the light-relic clustering is even less pronounced.
Therefore, it is safe for us to use the BKT approximation~\cite{Brandenberger:1987kf}.
In Appendix~\ref{app:Clustering} we find that the effect from clustering of light relics is largely scale-independent, and thus unimportant for our purposes.
Nevertheless, we will include it in our analysis unless otherwise stated.

\subsection{Lagrangian Bias}

Beyond their transfer function, which we calculate with {\tt CLASS},
light relics enter our calculation of the spherical collapse in two ways,
(\emph i) they modify the background cosmology, as in Eq.~\eqref{eq:Hubble}, and 
(\emph{ii}) they respond to long-wavelength CDM+b fluctuations, as in the right-hand side of Eq.~\eqref{eq:Rdotdot_Clust}.
In all the cases we consider in this work the new light component is nonrelativistic today, and thus contributes to the total matter energy density. 
To account for this, we will reduce the CDM density $\Omega_d$ by the necessary value to keep the total matter density $\Omega_m$ today fixed (we will always, of course, keep $\Omega_b$ fixed).
Additionally, we choose $z_{\rm coll}=0.7$, in line with the median redshift of galaxies observed in the dark energy survey (DES)~\cite{Bonnett:2015pww}.
For reference, at this redshift the nonlinear scale, $k_{\rm NL}$ (defined by demanding that the power per unit $\log(k)$ is unity, i.e., $
P_{mm}(k_{\rm NL}) k_{\rm NL}^3/(2\pi^2) = 1$),
is $k_{\rm NL} = 0.43\, h$ Mpc$^{-1}$ for our fiducial cosmology, although non-linear effects might start appearing at lower wavenumbers.
This scale roughly coincides with the comoving radius of the initial overdensities that collapse to form the haloes, as $k_* = \pi/R_M = 0.47\,h$ Mpc$^{-1}$ for the $M=10^{13}\,h^{-1} \, M_\odot$ haloes we consider, so we expect the bias to strongly depart from our linear predictions around that scale~\cite{Modi:2016dah}.
Nonetheless, for illustration purposes we will plot results up to $k \approx 1\, h$ Mpc$^{-1}$, to better show the behavior of different linear quantities beyond the free-streaming scale of the light relics.

\begin{figure}[hbtp!]
	\includegraphics[width=0.44\textwidth]{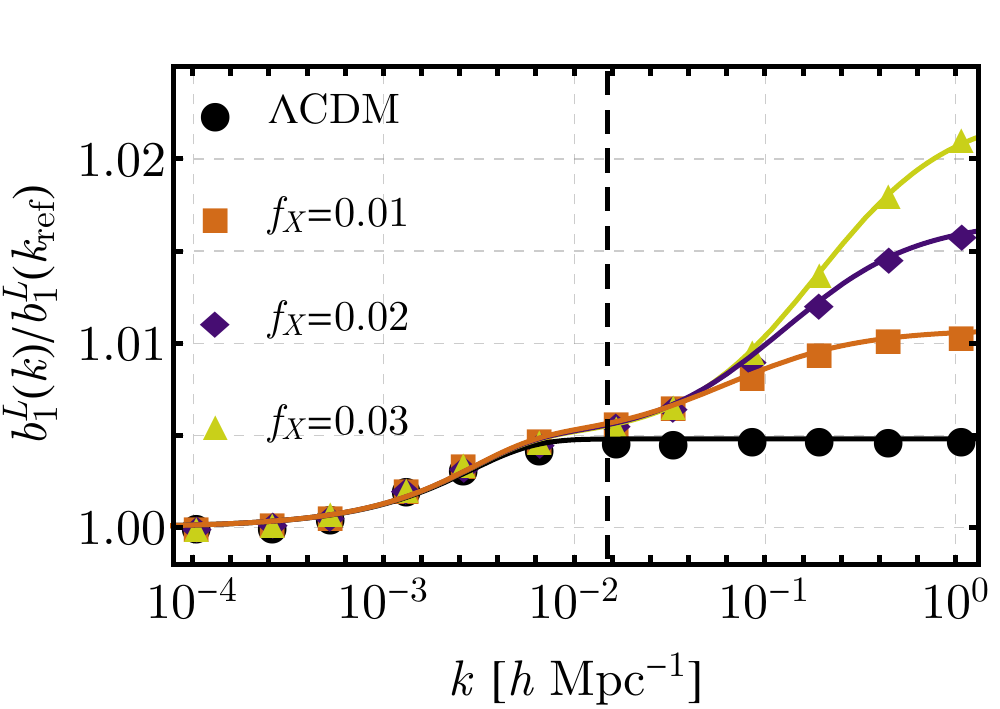}			
	\caption{
		Lagrangian bias obtained with {\tt RelicFast}, normalized at $k_{\rm ref}=10^{-4}\,h$ Mpc$^{-1}$, for haloes of mass $M= 10^{13} \,h^{-1}\, M_\odot$ collapsing at redshift $z_{\rm coll}=0.7$.
		We fix the light-relic temperature at $T_X^{(0)}=1.4$ K, to saturate the bound $\Delta N_{\rm eff}=0.3$, and change the fraction $f_X$ of matter in light relics. The three cases of $f_X=0.01, 0.02$ and $0.03$ correspond to particles with masses $m_X=0.35, 0.7,$ and $1.05$ eV.
		Solid lines show the fit from Eq.~\eqref{eq:fitbL} for each case, and 
		the vertical dashed line denotes the scale of matter-radiation equality.
	}
	\label{fig:bvsk_TR_Neff_Lag}
\end{figure}

We show in Fig.~\ref{fig:bvsk_TR_Neff_Lag} the Lagrangian bias for $\Lambda$CDM (with massless neutrinos) and for three light-relic cases, composing fractions $f_X=\{1\%, 2\%, 3\%\}$ of the total matter, all of which have a temperature $T_X^{(0)}=1.4$ K, chosen to saturate the 1-$\sigma$ $N_{\rm eff}$ bound from BBN.
These relics have masses of $m_X=\{0.35, 0.7, 1.05\}$ eV, so they would appear as a $\Delta N_{\rm eff}<0.3$ on the CMB, given that they transition to become nonrelativistic around the epoch of recombination.
Each bias is normalized with respect to its value at a reference wavenumber $k_{\rm ref}=10^{-4}\,h$ Mpc$^{-1}$.

For ease of visualization and understanding, we will provide a fit for the Lagrangian bias.
As noted before, $\Lambda$CDM shows a small difference in the growth of perturbations that reentered the horizon before and after matter-radiation equality, even in the absence of light relics (or massive neutrinos).
In order to include this in our analysis we use a simple step function as a fit,
\ba
\label{eq:RL_LCDM}
R_{L}^{\Lambda\rm CDM} \equiv& \dfrac{b_1^{L, \Lambda\rm CDM}(k)}{b_1^{L, \Lambda\rm CDM}(k_{\rm ref})}  \\ \nonumber =& 1 + \Delta_{\Lambda\rm CDM} \tanh\left( \alpha k/k_{\rm eq}\right),
\end{align}
where at $z_{\rm coll}=0.7$ we find $\alpha=4$, $\Delta_{\Lambda\rm CDM}=4.8\times 10^{-3}$, and $k_{\rm eq} = 0.015\,h$ Mpc$^{-1}$ is the scale of matter-radiation equality~\cite{Ade:2015xua}.

Light relics can cause a significant change in the bias at smaller scales, as seen in Fig.~\ref{fig:bvsk_TR_Neff_Lag}.
The size of the bias grows with $f_X$, and the scale at which it becomes important depends on the free-streaming scale of the light relic.
Given that the shape resembles a step function in $\log(k)$ space, we choose to fit it as
\be
\dfrac{b_1^{L,\rm fit}(k)}{b_1^L(k_{\rm ref})} = R_{L}^{\Lambda\rm CDM}   \left [1 + \dfrac{\Delta_L}{2} \left(\tanh\left[\dfrac{\log(q)}{\Delta_q}\right]+1\right) \right],
\label{eq:fitbL}
\ee
where we find $\Delta_L=0.6 \, f_X$,  $q \equiv 5\,k/k_{\rm fs}$, and $\Delta_q=1.6$.
A more precise fit can, of course, be achieved, at the cost of making the fitting function more complicated.
Nonetheless, we will see that this simple functional form provides a reasonably good fit to all cases we will study.
Note that we have decided to employ $k_{\rm fs}$ to parametrize the scale at which the step arises in the bias. Using $k_{\rm fsh}$ or $k_{\rm nr}$ would be equivalent, as these quantities are linearly related, barring a mild redshift dependence.
We have checked that this function provides an excellent fit for other halo masses, and is not altered significantly for other redshifts.
This shows that, even though the overall value of the Lagrangian bias is strongly dependent on redshift and halo mass, the scale dependence induced by light relics is not. We will elaborate on this later.

\subsection{Eulerian Bias}

\begin{figure}[hbtp!]
	\includegraphics[width=0.44\textwidth]{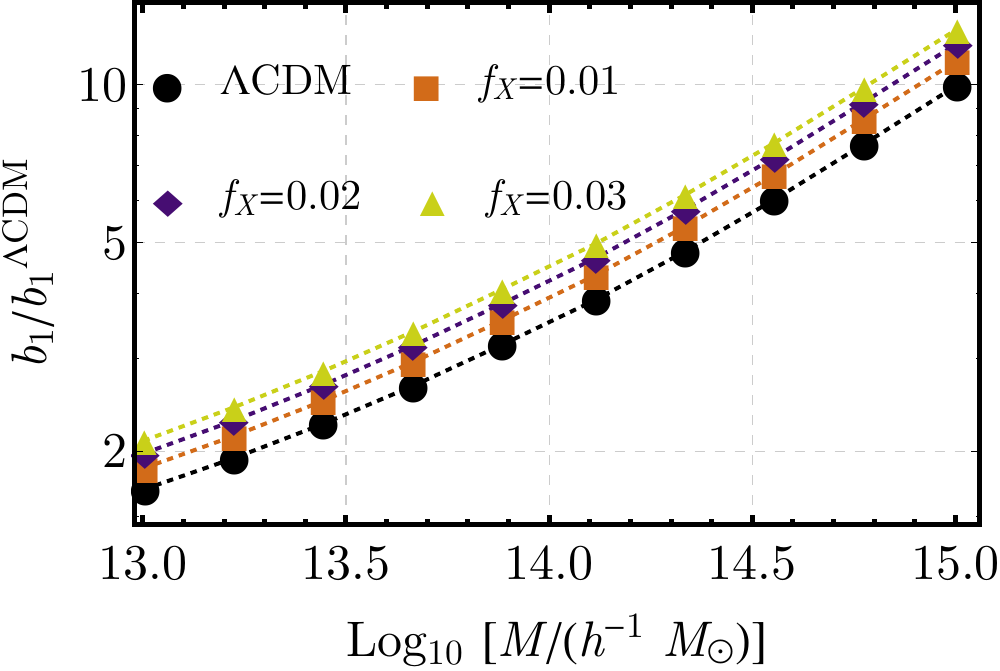}			
	\caption{Linear (Eulerian) bias at $z_{\rm coll}=0.7$ and $k_{\rm ref}= 10^{-4}\,h$ Mpc$^{-1}$, as a function of halo mass, for the same light-relic cases as in Fig.~\ref{fig:bvsk_TR_Neff_Lag}.
	Dotted lines join the points to guide the eye.
	}
	\label{fig:bvsM_TR}
\end{figure}

We show in Fig.~\ref{fig:bvsM_TR} the linear Eulerian bias, at the redshift of collapse, as a function of halo mass for the three light-relic cases defined above, as well as the case of $\Lambda$CDM. We see that adding light relics causes haloes of all masses to be more biased, as we are substituting some CDM for (warmer) relics, and thus the growth of fluctuations is decreased. This change in the bias is, to a large extent, mass independent, so it might be difficult to observe.

\begin{figure}[hbtp!]
	\includegraphics[width=0.44\textwidth]{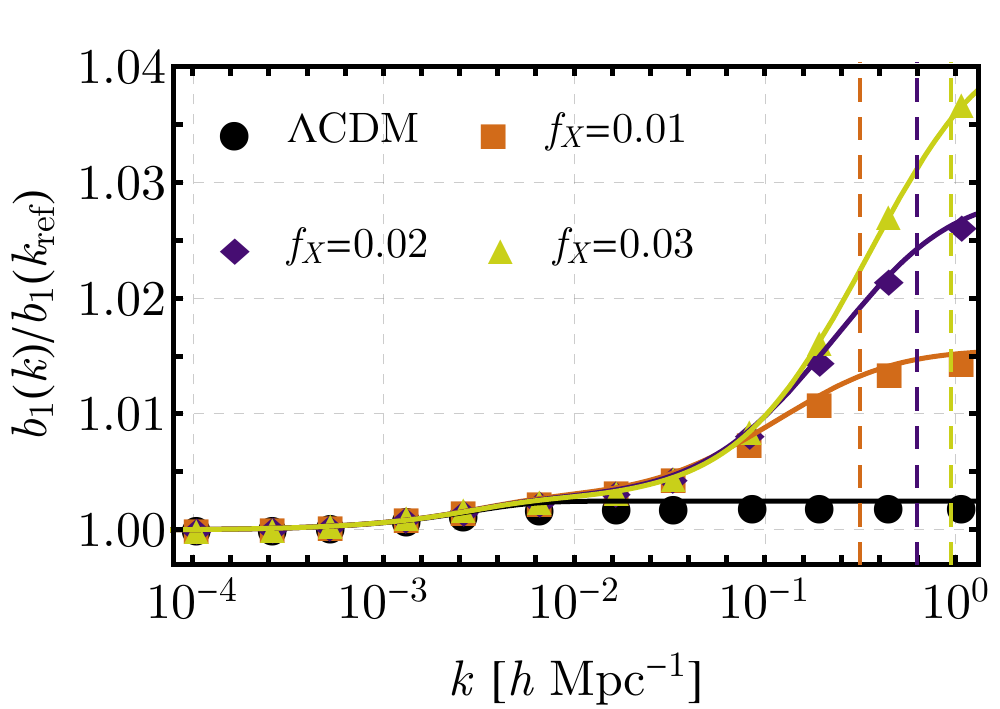}			
	\caption{
	We show the normalized linear Eulerian bias with the same inputs as in Fig.~\ref{fig:bvsk_TR_Neff_Lag}.
	Solid lines are obtained using Eq.~\eqref{eq:b1_fit}, with the fit for the Lagrangian bias of Eq.~\eqref{eq:fitbL} and the transfer functions from {\tt CLASS}.
	The vertical dashed lines show the free-streaming scale of each of the particles considered.
	}
	\label{fig:bvsk_TR_Neff}
\end{figure}

The (scale-independent) value of $b_1$ is usually marginalized over in galaxy surveys, as a host of complexities can affect it. Thus, we will focus on the scale-dependence of $b_1$, as we did for $b_1^L$.
We show $b_1$ as a function of wavenumber in Fig.~\ref{fig:bvsk_TR_Neff}.
We use the fit for $b^L_1 (k)$, from Eq.~\eqref{eq:fitbL}, to find the fitted Eulerian bias as
\be
b_1^{\rm fit} (k) = \left[1+b_1^{L,\rm fit} (k)\right] \dfrac{\mathcal T_c(k)}{\mathcal T_m(k)},
\label{eq:b1_fit}
\ee
which follows trivially from Eq.~\eqref{eq:b1}, where we have defined the matter transfer function
\be
\mathcal T_m(k) = (1-f_X) \mathcal T_c(k)+ f_X \mathcal T_X(k),
\ee
in terms of the CDM+b ($\mathcal T_c$) and light-relic ($\mathcal T_X$) ones, which are calculated with {\tt CLASS} at $z_{\rm coll}$.
From Figs.~\ref{fig:bvsk_TR_Neff_Lag} and \ref{fig:bvsk_TR_Neff} we also see that the scale-dependence of the biases starts at scales larger than the free-streaming scale of the light relics, as found in Ref.~\cite{LoVerde:2014pxa} for the case of massive neutrinos.

Our prediction for the scale dependence of the Lagrangian bias is a step-like function, with a plateau at $k \sim k_{\rm fs}$, which yields a step-like Eulerian bias as well.
Even though this is different from the expected $k^2$ scaling of the bias that appears at small scales in $\Lambda$CDM~\cite{Musso:2012qk,Modi:2016dah}, for large $k_{\rm fs}$ these two effects might be indistinguishable for all practical purposes, allowing searches of heavy relics without the need of solving for spherical collapse.
Additionally, part of the scale dependence of $b_1(k)$ can be attributed to the reduction of the matter power spectrum at small scales, as relic fluctuations vanish~\cite{Castorina:2013wga,Villaescusa-Navarro:2013pva}. 
However, even if one defined the bias with respect to CDM+b only, as opposed to all matter, there would still be some scale dependence~\cite{LoVerde:2014pxa}, arising from the behavior  of $b_1^L(k)$, which we showed in Fig.~\ref{fig:bvsk_TR_Neff_Lag}. We explore this question in Section~\ref{sec:Neutrinos} for the case of neutrinos.
Nonetheless, since the observable quantity is the halo overdensity, it makes little difference how we define the bias, as long as we are self consistent.

\begin{figure}[hbtp!]
	\includegraphics[width=0.49\textwidth]{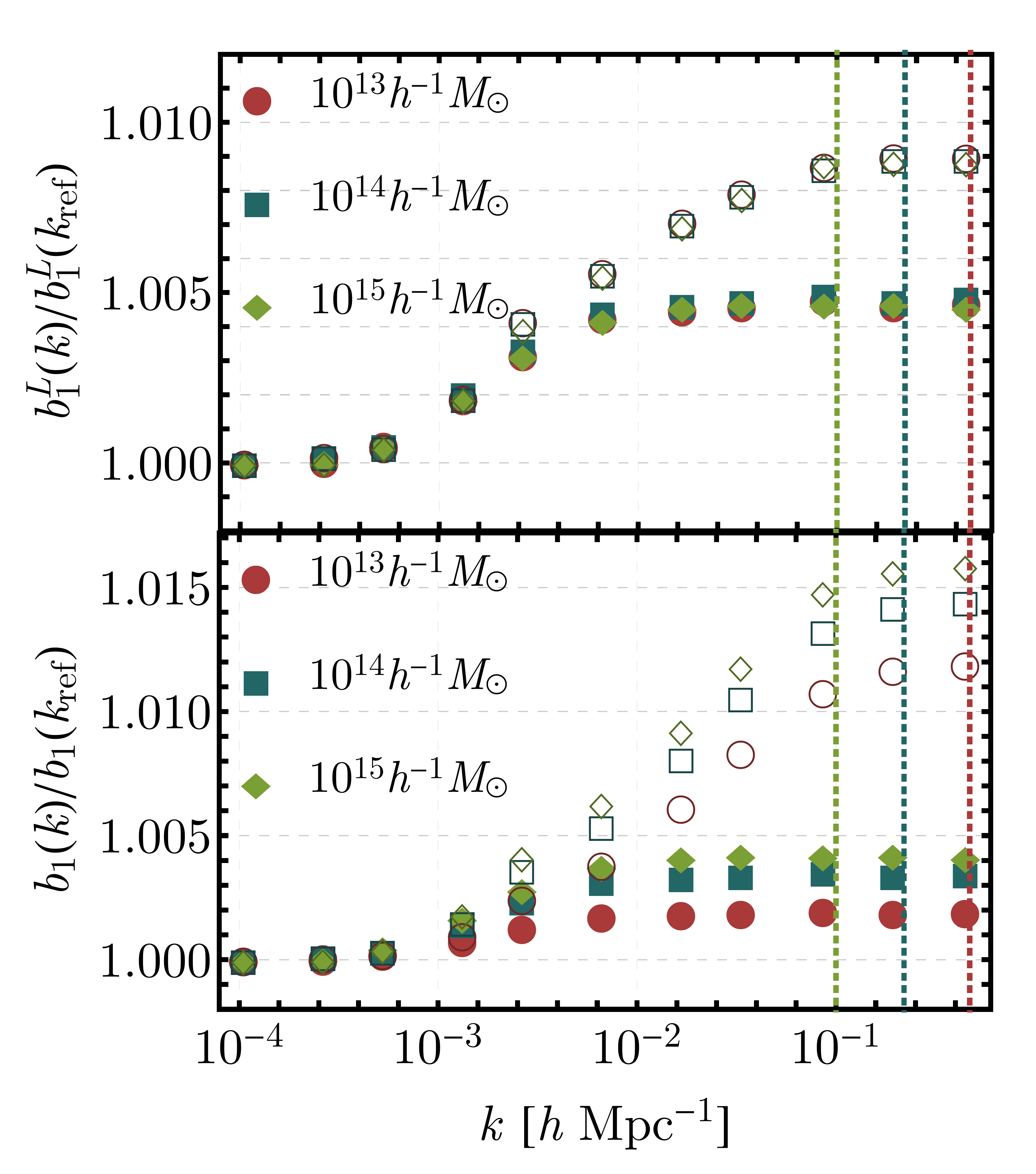}					
	\caption{Linear biases both for $\Lambda$CDM (in filled symbols), and for a cosmology with one massive neutrino (with $m_{\nu_1}=0.1$ eV; in hollow symbols), normalized at $k_{\rm ref}=10^{-4}\,h$ Mpc$^{-1}$.  We vary the halo mass $M$, assuming a redshift of collapse of $z_{\rm coll}=0.7$. We ignore neutrino clustering, and we keep the CDM density today fixed.
	We warn the reader that  haloes of masses $M=\{10^{13},10^{14},10^{15}\}\,h^{-1}\,M_\odot$ are formed from overdensities of typical comoving wavenumber $k_*=\{0.47,0.22,0.10\} \,h $ Mpc$^{-1}$, shown as vertical dotted lines,  beyond where the bias will  depart from our linear predictions.
	}
	\label{fig:bias_Mhalo}
\end{figure}

We want to point out that the scale dependence of $b_1$ is more susceptible to changes in the properties of the haloes than the scale dependence in $b_1^L$~\cite{LoVerde:2014pxa}.
To showcase this effect , we have calculated both the linear Eulerian and Lagrangian biases for a cosmology with one massive neutrino, of mass $m_{\nu_1}=0.1$ eV, as well as for $\Lambda$CDM, for three halo masses. We show the biases for these cases in Fig.~\ref{fig:bias_Mhalo}, from where we see that the normalized Lagrangian bias is nearly identical for all halo masses, whereas the normalized Eulerian bias shows a larger spread in values. 
This is not surprising, as the overall value of the Lagrangian bias, $b_1^L(k_{\rm ref})$, enters the calculation of $b_1$, and this quantity is very different for the three halo masses we show, with values of $b_1^L(k_{\rm ref})\approx 0.5$, 2, and 7. 
A similar effect arises when varying the collapse redshift, as we show in Appendix~\ref{app:Code}, albeit it is less pronounced.
Given this insight, and the additional myriad of effects that can affect the overall amplitude of the Lagrangian bias, we encourage users to marginalize over the amplitude of the Lagrangian bias, $b_1^L(k_{\rm ref})$, as opposed to that of the Eulerian one, to keep the scale dependence in the most pristine state.

\subsection{Power Spectrum}

\begin{figure}[hbtp!]
	\includegraphics[width=0.44\textwidth]{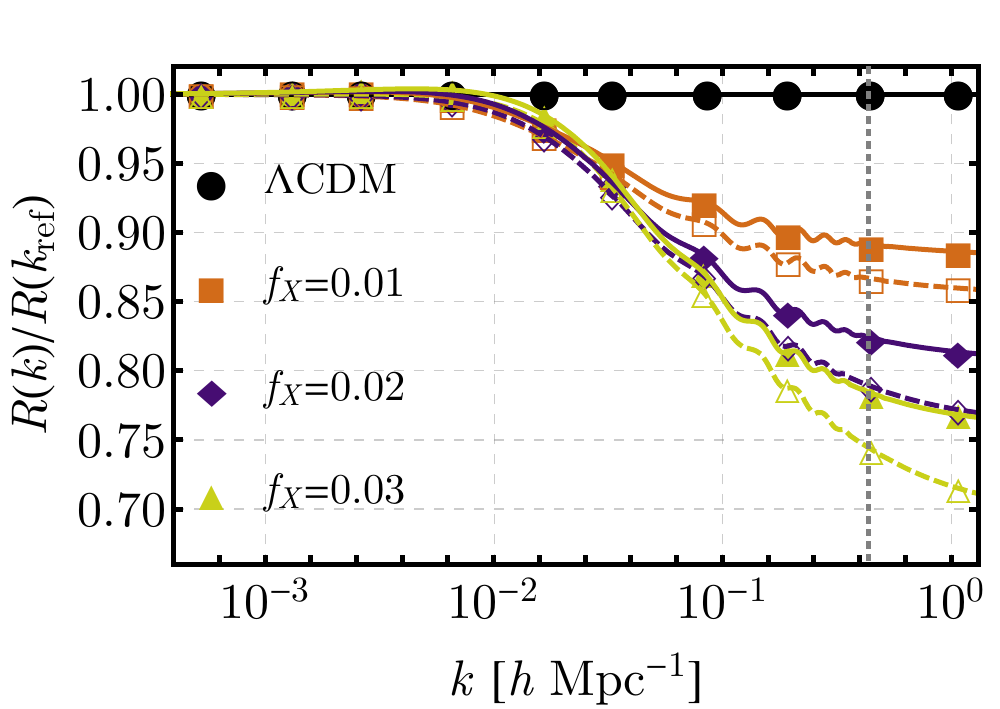}			
	\caption{
	Normalized suppression factors, as defined in Eq.~\eqref{eq:transfer}, for the same parameters as Fig.~\ref{fig:bvsk_TR_Neff_Lag}.
	We show in hollow symbols the suppression $R_m$ of the matter power spectrum in the presence of light relics, and in filled symbols that of the halo power spectrum, $R_h$.
	Dashed lines represent $R_m$, and are obtained with {\tt CLASS} output, whereas solid lines represent $R_h$, and include our fit for the bias from Eq.~\eqref{eq:b1_fit}.
	The vertical grey-dotted line represents the nonlinear scale $k_{\rm NL}$ at $z=0.7$.
	}
	\label{fig:Supp_TR_Neff}
\end{figure}

Cosmological relics suppress the matter power spectrum, as we discussed in Section~\ref{sec:Relics}.
In order to quantify this effect,
let us define the suppression factor
\be
R_s(k) \equiv  \dfrac{P_{ss}(k)}{P_{ss}^{\Lambda\rm CDM}(k)} ,
\label{eq:transfer}
\ee
where the index $s=\{m,h\}$ stands for matter or halo power spectra. 
We work in real space (as opposed to redshift space), so we can relate the halo and matter power spectra simply as
\be
P_{hh} (k) = b_1^2(k)  P_{mm}(k),
\ee
from where we can easily calculate both $R_m$ and $R_h$ as a function of scale.
We show these two quantities, normalized at large scales, in Fig.~\ref{fig:Supp_TR_Neff}.
We find that the matter power spectrum is suppressed for scales beyond $k\sim 10^{-2} \, h$ Mpc$^{-1}$, asymptoting to a value of $R_m(k)/R_m(k_{\rm ref}) \approx (1-14 f_X)$ at very small scales, as found in Ref.~\cite{Boyarsky:2008xj}.
However, the scale dependence of the bias reduces the suppression by tens of percent, making the effect of light relics less obvious in galaxy data.

Note that the lines with light relics in Fig.~\ref{fig:Supp_TR_Neff} show wiggles when compared to $\Lambda$CDM.
In addition to the well-known shift in the BAO phase caused by the addition of free-streaming particles~\cite{Bashinsky:2003tk},
we are changing the background cosmology when adding relics, as part of the matter density will become relativistic at high-enough redshift. 
The relics in Fig.~\ref{fig:Supp_TR_Neff} become nonrelativistic roughly at $z\sim 3\times 10^3-10^4$, which changes the sound horizon at recombination with respect to a universe with the same amount of matter today, but no light relics.
Heavier relics become nonrelativistic earlier, and thus the wiggles are less pronounced for larger $f_X$ in Fig.~\ref{fig:Supp_TR_Neff}.
In any case, part or all of this effect can be reabsorbed, for instance, in the inferred value of $h$, to obtain the same sound horizon.
Additionally, since the relics we study are nonrelativistic during recombination, their effect on the power spectrum appears at smaller scales than $k_{\rm eq}$. This might not be a disadvantage, as galaxy surveys, which often cannot observe modes longer than $k\sim10^{-2} \, h$ Mpc$^{-1}$, would be sensitive to the ``turning on" of the suppression, perhaps making these relics easier to constrain than neutrinos.

\section{Other Relics}
\label{sec:Other_Relics}

So far we have, for simplicity, only considered the case of a neutrino-like (spin-1/2 single-family) relic. 
Nonetheless, we will now show that a host of other light relics can be expressed as an \emph{equivalent} neutrino in terms of their cosmological effects.
We will use this to show results for other fermions, focusing on non-resonantly produced (NRP) sterile neutrinos. We will also show how bosonic degrees of freedom can be approximately represented as an equivalent neutrino as well, and exemplify this with scalar and vector light relics.

\subsection{Other Fermions}

At the level of perturbations, it was shown in Ref.~\cite{Boyarsky:2008xj} that a non-resonantly produced (NRP) neutrino is equivalent to a regular neutrino, with an appropriate choice of mass and temperature. 
In general, we expect a fermionic particle, with a number $g_Y$ of degrees of freedom, a mass $m_Y$, and a temperature $T_Y$, to be equivalent to the one-family neutrino case we have studied ($X$) if the following relations are satisfied:
\begin{subequations}
\ba
\Omega_Y &= \Omega_X, \quad {\rm and} \\
\dfrac{T_Y^{(0)}}{m_Y} &= \dfrac{T_X^{(0)}}{m_X}.
\end{align}
\end{subequations}
We can solve these equations to find the equivalent-neutrino mass and temperature as
\begin{subequations}
\ba
m_X &= m_Y (g_Y/g_X)^{1/4}, \quad {\rm and} \\
T_X^{(0)} &= T_Y^{(0)} (g_Y/g_X)^{1/4},
\label{eq:YtoX}
\end{align}
\end{subequations}
where $g_X=3/2$.
We note that two light relics following this relation would also contribute with the same $\Delta N_{\rm eff}$ at any epoch, and thus either both satisfy, or violate, CMB and BBN bounds.
It is in this sense that we call these relics ``equivalent", as all their background and linear-level cosmological effects are identical.

\begin{figure}[hbtp!]
	\includegraphics[width=0.44\textwidth]{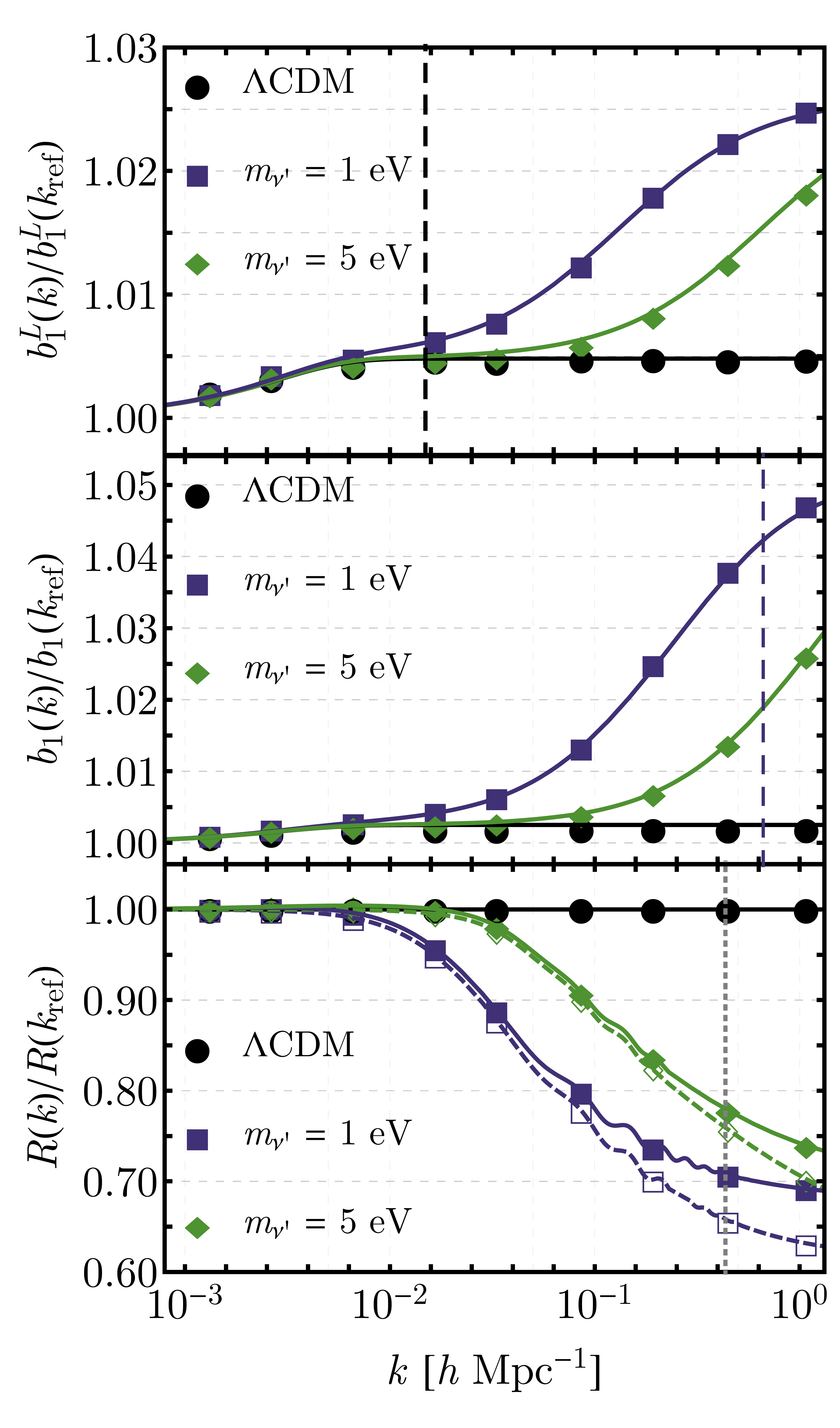}
	\caption{
		Lagrangian and Eulerian biases, and suppression factors, for the case of sterile neutrinos of different masses, composing 3.5\% of the total matter (and thus saturating the Planck bound). 
		The amplitudes of their distributions are $\chi=1/2$ and $1/10$ for $m_{\nu'}=1$, and $5$ eV, respectively.
		We consider a halo with $M=10^{13}\,h^{-1}\,M_\odot$, $z_{\rm coll}=0.7$, and normalize all results at $k_{\rm ref}= 10^{-4}\, h\, \rm Mpc^{-1}$.
		As before, solid lines represent our fit for the biases in the top two panels, and in the bottom panel solid lines and filled symbols represent suppression in halo power spectra, whereas dashed lines and empty symbols correspond to matter power spectra.
	}
	\label{fig:TR_sterile}
\end{figure}

We will use this equivalence to study the case of NRP sterile neutrinos ($\nu'$) composing a fraction of the cosmological dark matter~\cite{Dodelson:1993je,Gariazzo:2015rra}.
In particular, sterile neutrinos with eV masses have received wide interest, as they could explain some observed short-baseline neutrino anomalies (see for instance Refs.~\cite{Dentler:2018sju,Capozzi:2018ubv} for recent analyses).
These particles cannot compose all of the DM, as they are too light to be CDM, but
their number density can be suppressed by many processes, such as interactions with a dark photon~\cite{Dasgupta:2013zpn}, yielding small cosmic abundances.
Therefore, we assume that the NRP neutrinos have a modified Fermi-Dirac distribution given by
\be
f^{\rm NRP}_{\nu'}(q,z) = \dfrac{\chi}{e^{\,q/T_\nu(z)}+1},
\label{eq:fNRP}
\ee
where $\chi\leq 1$ is a normalization factor, chosen to set the right NRP neutrino abundance, given its mass $m_{\nu'}$~\cite{Boyarsky:2008xj}.
Sterile neutrinos share the active neutrino temperature ($T_{\nu'}^{(0)}=T_\nu^{(0)}$).
Interestingly, for small enough $\chi$ the temperature $T_X^{(0)}$ of the equivalent light relic, calculated with Eq.~\eqref{eq:YtoX} and $g_{\nu'} = 3 \chi /2$, can be below $T_X^{(0)}<1$ K---the lowest standard-model prediction.

Our analysis is similar to that of Ref.~\cite{Ade:2015xua}, where instead of varying $N_{\rm eff}$ and the ``effective" neutrino mass, defined as
\be
m_{\nu'}^{\rm eff} = 94.1 \, {\rm eV} \times \Omega_{\nu'} h^2,
\ee
we vary $m_{\nu'}$ and $\Omega_{\nu'}$, as those parameters are more relevant for galaxy observables. 
The Planck collaboration found the 2-$\sigma$ constraint $m_{\nu'}^{\rm eff} < 0.5$ eV~\cite{Ade:2015xua}, 
which can be translated into a fraction $f_{\nu'}<0.035$ of the total matter on sterile neutrinos.
We will keep this value fixed, and vary the mass of the sterile neutrino, for which we will simply rescale the $\chi$ factor in Eq.~\eqref{eq:fNRP}.
We show two cases in Fig.~\ref{fig:TR_sterile}, corresponding to $m_{\nu'}=1$ and $5$ eV (or $\chi=1/2$ and $1/10$).
To obtain these results, we have ran {\tt RelicFast} with $m_X=0.84$ eV and 2.8 eV, taking $\Omega_X h^2=5\times10^{-3}$ in both cases, which corresponds to temperatures $T_X^{(0)}=1.7$ K and 1.1 K for the equivalent neutrino.
Note, in passing, that these two relics become nonrelativistic at $z\approx 6\times 10^3$ and $3\times 10^4$, respectively, so they would not fully contribute to the radiation density during the CMB epoch. In addition, these relics produce $\Delta N_{\rm eff} = 0.5$ and $0.1$ at BBN, within current 2-$\sigma$ limits~\cite{Cyburt:2015mya}.

\begin{figure}[h!]
	\includegraphics[width=0.44\textwidth]{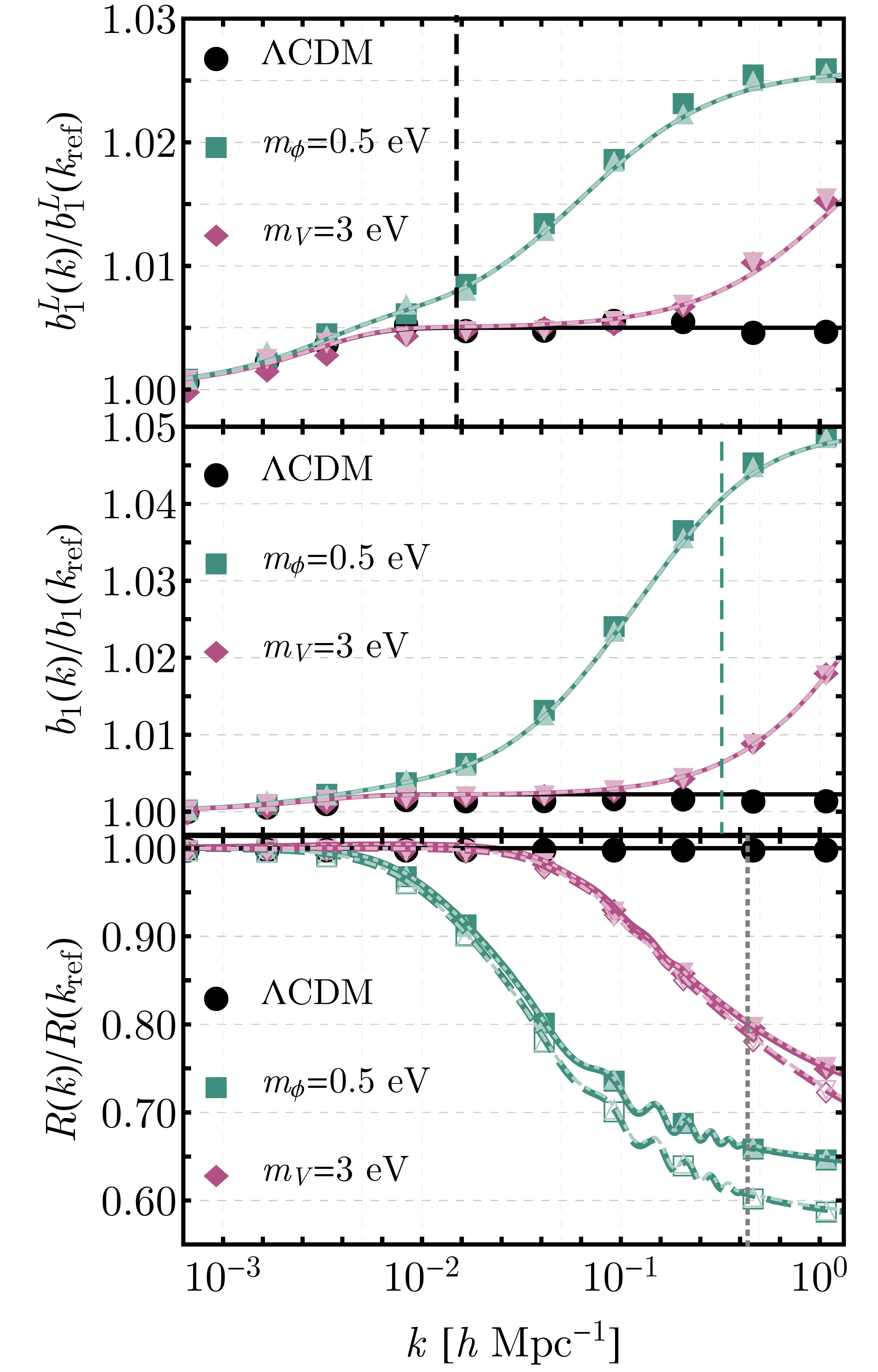}
	\caption{
		Same as Fig.~\ref{fig:TR_sterile}, but for the case of two bosons, and their equivalent neutrinos, all of them composing 3.5\% of the matter today. 
		In (dark) green we show the case of a scalar of mass $m_\phi = 0.50$ eV, and  in light green its corresponding equivalent neutrino of mass $m_X = 0.51$ eV. Similarly, the case of a vector of mass $m_V=3.0$ eV is shown in (dark) pink, where the equivalent neutrino (in light pink) has $m_X=4.0$ eV. As before, lines are obtained with the fits of Eqs.~(\ref{eq:fitbL},\ref{eq:b1_fit}) and the transfer functions from {\tt CLASS}, with dashed lines (and hollow symbols) representing $R_m$, and solid lines (with filled symbols) representing $R_h$.
	}
	\label{fig:TR_boson}
\end{figure}

\subsection{Bosons}

Bosonic relics, such as scalars and vectors, cannot be trivially expressed in terms of a fermion, given that their momenta are distributed according to a Bose-Einstein distribution, instead of a Fermi-Dirac one.
These two distributions, however, share the same ultrarelativistic and nonrelativistic limits.
Therefore, we can approximate bosonic ($Y$) and fermionic ($X$) relics, by demanding
\begin{subequations}
	\ba
	\Omega_Y &= \Omega_X, \quad {\rm and} \\
	\VEV{v_Y}_{\rm BE} &= \VEV{v_X}_{\rm FD}\! ,
\end{align}
\end{subequations}
where $\VEV{}_{\rm FD}$ and $\VEV{}_{\rm BE}$ mean average under a Fermi-Dirac or Bose-Einstein distribution\footnote{It might seem more appropriate to demand $\VEV{v^{-2}}$ to be the same for the two distributions, instead of $\VEV{v}$, given the definition of $k_{\rm fs}$ in Eq.~\eqref{eq:q:kfs}. Nonetheless, this quantity is ill-defined for a BE distribution, and we will see that our definition suffices to make bosonic and fermionic relics indistinguishable.}.
The average velocity  of fermionic particles is 
$\VEV{v_X}_{\rm FD}= c_{\rm FD} \times T_X/m_X$, with $c_{\rm FD}\approx 3.15$, whereas for bosons it is $\VEV{v_Y}_{\rm BE}= c_{\rm BE} \times T_Y/m_Y$, with $c_{\rm BE}\approx 2.70$, where the ratio $c_{\rm FD}/c_{\rm BE} = 7/6$ exactly.
Moreover, the nonrelativistic FD and BE energy densities can be expressed as $\Omega_{(X/Y)} = g_{(X/Y)} m_{(X/Y)} T^3_{(X/Y)} C$,
where the $C$ factor is cosmology-dependent and common for both cases.
Thus, we can relate the bosonic and fermionic degrees of freedom by
\begin{subequations}
	\ba
	m_X &= m_Y (g_Y/g_X)^{1/4}  (7/6)^{3/4} \quad {\rm and} \\
	T_X &= T_Y (g_Y/g_X)^{1/4}(6/7)^{1/4},
	\label{eq:YtoX_boson}
\end{align}
\end{subequations}
where we remind the reader that each bosonic degree of freedom contributes with $g_Y=1$, and each fermionic one with $g_X=3/4$.
Then, we can find the result for an arbitrary particle with any spin and mass in terms of our one-family spin-1/2 case, where $g_X = 3/2$.

For illustrative purposes we will consider two different bosonic cases, a scalar with mass $m_\phi=0.5$ eV, and a vector with mass $m_V=3$ eV, both with an energy density today of $\Omega_Y h^2 = 5 \times 10^{-3}$, i.e., composing $3.5\%$ of the total matter.
To find the result for bosons,  we have modified the {\tt CLASS} code to allow for bosonic light relics, and have substituted the FD distribution in our code for a BE. 
Since the fluid nature of bosonic relics is uncertain, we choose to set their sound speed in our analysis to $c_s^2=w$ (and we do the same for the equivalent neutrinos, to allow for a ready comparison).
We show the biases, as well as the suppression factors for these two cases in Fig.~\ref{fig:TR_boson},
where we ignore clustering of these degrees of freedom, as the usual formulas in Appendix~\ref{app:Clustering} are only valid for fermions.

To test the validity of the transformations in Eq.~\eqref{eq:YtoX_boson},
we compare the results for bosons with the equivalent-neutrino approximation outlined above, for which the scalar and vector cases correspond to $m_X=0.51$ eV, and $m_X=4.0$ eV, respectively, both with $\Omega_X h^2=5\times 10^{-3}$.
We see that the two results are indistinguishable both in terms of bias and power spectra, as the  equivalent-neutrino approximation holds excellently well.
We note that even the absolute bias is the same under this approximation, at the 0.1\% level.
While this means that we can efficiently express any relic, even bosons, as an equivalent neutrino, it also means that elucidating the spin of a relic is virtually impossible.

\section{Neutrinos}
\label{sec:Neutrinos}

Perhaps the best-studied case of light relics is that of neutrinos.
Neutrinos are certain to populate our Universe, composing almost half of the energy density in the radiation-dominated plasma after BBN.
Interestingly, neutrino-oscillation experiments have shown that at least two of the three propagation eigenstates are massive, with mass-squared differences of $m_2^2 - m_1^2 =  (9 \,\rm meV)^2$ and $|m_3^2 - m_1^2| = (50 \,\rm meV)^2$~\cite{Fogli:2012ua,Abazajian:2013oma}.
However, the ``zero-point" of these masses is not known, and neither is the sign of $m_3-m_1$. Therefore, given a sum of neutrino masses, two hierarchies can be assumed: the normal hierarchy (NH), where $m_3>m_2\sim m_1$, and the inverted hierarchy (IH), where $m_2\sim m_1>m_3$.
Finding the sum of the neutrino masses, as well as which hierarchy is represented in nature, is a goal of present-day cosmology, and may well be within the reach of upcoming observations.

As an example of the sensitivity of current measurements, Planck data alone can constrain the sum of neutrino masses to be $\sum_i m_{\nu_i} < 0.49$ eV, at 95\%~C.L.~\cite{Ade:2015xua}, which can be further tightened to $\sum_i m_{\nu_i} < 0.17$ eV when adding distance information from BAO surveys~\cite{Beutler:2011hx,Anderson:2013zyy,Ross:2014qpa}.
This limit is, nonetheless, loosened to $\sum_i m_{\nu_i} < 0.29$ eV when adding the power spectra observed by the DES, as these probes are in mild tension with each other~\cite{Abbott:2017wau}.
This illustrates that in order to find a definitive measurement of neutrino masses, we have to study the effect of neutrinos beyond the background cosmology.

Most neutrino-mass searches with cosmological data make two simplifying approximations. First, they vary the sum of the neutrino masses, $M_\nu = \sum_i m_{\nu_i}$, either assuming that all neutrinos have the same mass, or that only one neutrino is massive.
This has been shown to be a good approximation within the precision of current data~\cite{Lesgourgues:2004ps,Jimenez:2010ev}, but might not be true with next-generation surveys.
Second, these searches commonly assume a scale-independent bias, over which they marginalize, ignoring the effect of neutrinos in the halo bias~\cite{Vagnozzi:2017ovm,dePutter:2012sh}.
Recently, this assumption has been relaxed in Ref.~\cite{Giusarma:2018jei}, where a scale-dependent bias of $k^2$ form was included, which arises naturally in $\Lambda$CDM, but is unrelated to the scale dependence induced by neutrinos.

\begin{figure}[h!]
	\includegraphics[width=0.44\textwidth]{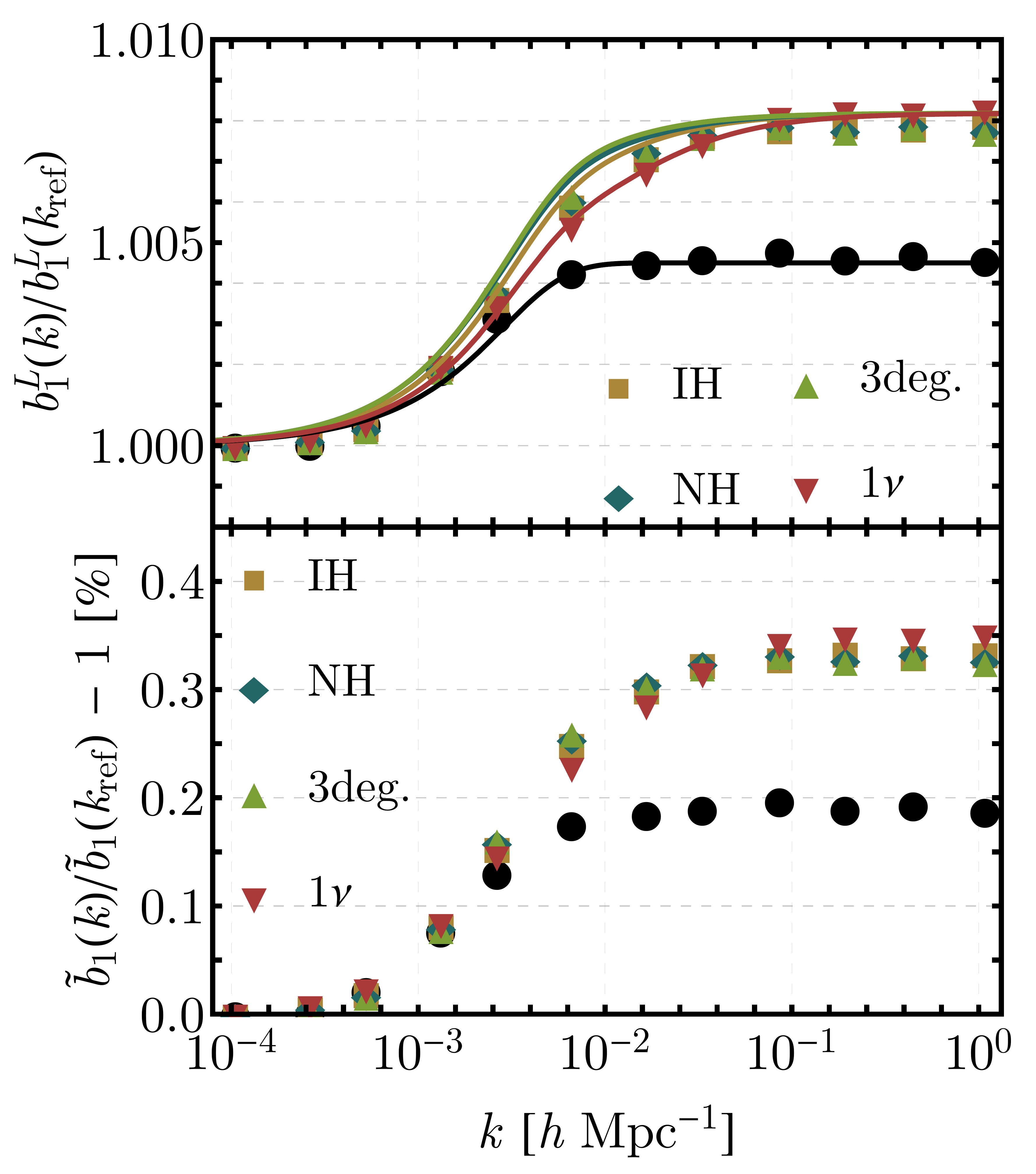}	
	\caption{
		Linear Lagrangian bias and Eulerian bias with respect to CDM (defined as $\tilde b_1 = P_{hc}/P_{cc}$), for the IH, NH, and the approximations with only one neutrino (1$\nu$) and three degenerate neutrinos (3deg.), all with same total mass $\sum m_{\nu_i} = 0.09$ eV, as well as for $\Lambda$CDM with massless neutrinos (in black).
		The fit for the bias is obtained with Eq.~\eqref{eq:fitbL_nus}.
		Similarly to other figures,
		we set $M=10^{13}\,h^{-1}\,M_\odot$, $z_{\rm coll}=0.7$, and $k_{\rm ref}= 10^{-4}\, h\, \rm Mpc^{-1}$, although we keep $\Omega_d$ fixed instead of $\Omega_m$.
	}
	\label{fig:bias_nus_cdm}
\end{figure}

We address these two issues with {\tt RelicFast}. We obtain the transfer functions from the Boltzmann solver {\tt CLASS}, which allows for any number of light relics with different masses (and temperatures)~\cite{Lesgourgues:2011rh,Brinckmann:2018cvx}.
Moreover, we implement the spherical collapse of haloes including all neutrinos simultaneously, which provides us with the halo bias in the presence of three neutrinos with arbitrary masses.
Then, we find the halo power spectrum for both the NH and IH, for any $M_\nu$, and ask whether considering one massive neutrino (1$\nu$)  or three degenerate ones (3deg.)  are good approximations to either of the two hierarchies.

Throughout this section we will keep the CDM density $\Omega_d h^2=0.12$ fixed, as a proxy for CMB observations, since light neutrinos would always appear as radiation during recombination (see, however, Ref.~\cite{Archidiacono:2016lnv} for the impact of CMB lensing).
This means that $\Omega_m h^2$ will be larger for universes with massive neutrinos, and thus $\Omega_\Lambda$ will be reduced (as we do not alter $h$).
We show in Appendix~\ref{app:Omegam} how keeping  $\Omega_m$ fixed, instead of $\Omega_c$, produces a larger suppression in the power spectra but almost identical biases.
Additionally, we have not included the effects of neutrino clustering since it has been shown to be negligible for the neutrino masses we study here~\cite{LoVerde:2014rxa}.
Finally, we reduce $N_{\rm eff}$ by $N_{\rm eff}^{\nu} = 1.0132$ for each massive neutrino we independently include, both in our calculation and in {\tt CLASS}, so as to produce $N_{\rm eff}=3.046$ at early times~\cite{Lesgourgues:2011rh}.

\begin{figure}[h!]
	\includegraphics[width=0.44\textwidth]{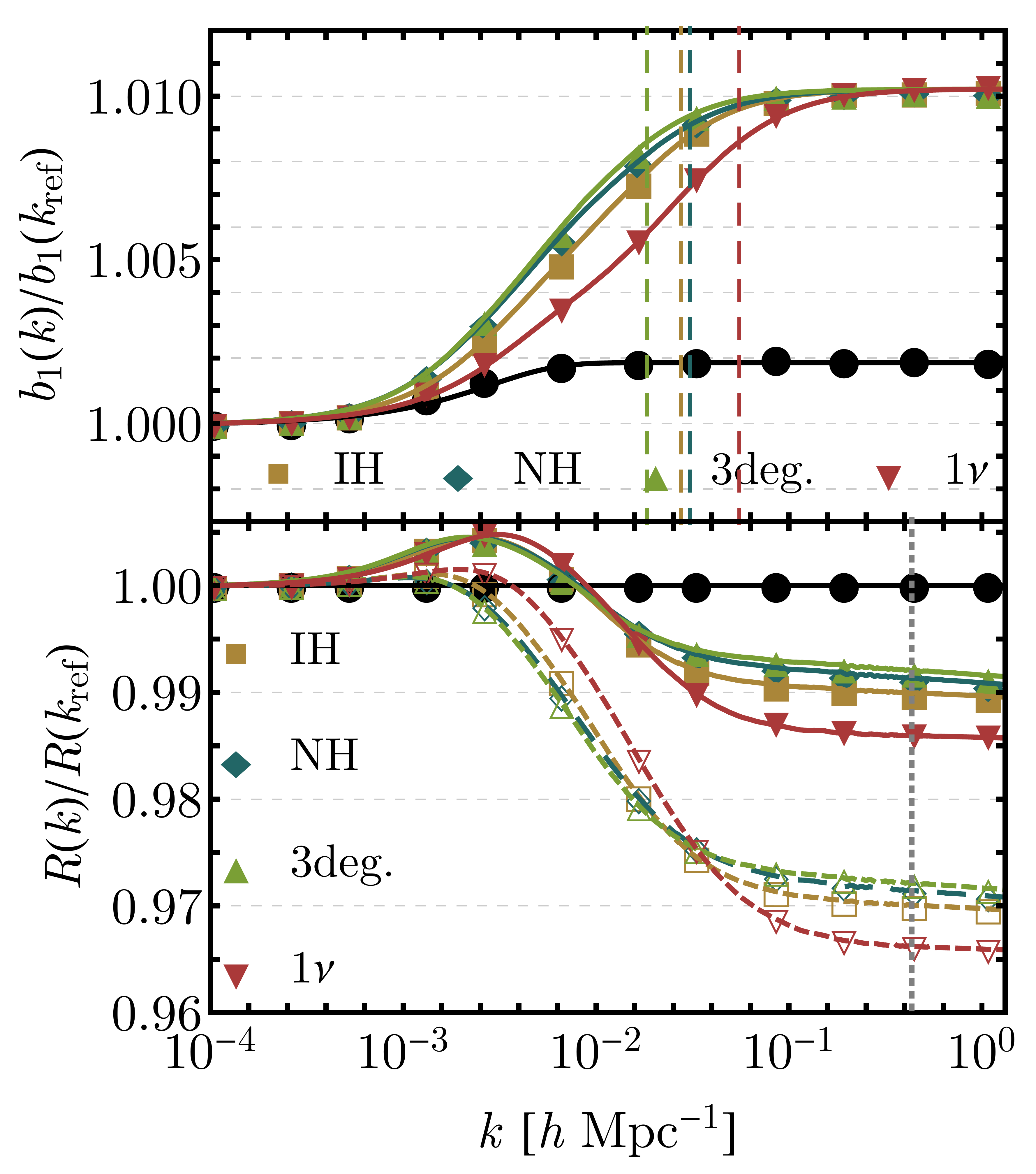}		
	\caption{
		Linear bias and suppression factors for the same inputs as Fig.~\ref{fig:bias_nus_cdm} (i.e., $\sum_i m_{\nu_i} =0.09$ eV).
		The vertical dashed lines represent the largest $k_{\rm fs}$ for each case, and
		in hollow symbols and dashed lines in the bottom panel denote $R_m$, whereas filled symbols and solid lines denote $R_h$.
	}
	\label{fig:Supp_9eV}
\end{figure}

\subsection{Scale-dependent Bias}

We begin by finding the effect of neutrinos on the linear bias.
We will treat each massive neutrino as an independent light relic. Thus, we will consider an arbitrary number $N_\nu$ of them, by self-consistently including them both in the Boltzmann solver ({\tt CLASS}) and in the spherical collapse equation.
We show the result for the Lagrangian bias in Fig.~\ref{fig:bias_nus_cdm}, for both the NH and IH, as well as the 1$\nu$ and 3deg.~approximations.
From this Figure we see that, even with a modest total mass of 0.09 eV, neutrinos cause a  0.35 percent step in the Lagrangian bias, in addition to the half a percent already present in $\Lambda$CDM.

In order to approximate our result, we fit the combined effect of a number $N_\nu$ of massive neutrinos through
\be
\dfrac{b_1^{L,\rm fit}(k)}{b_1^L(k_{\rm ref})} = R_{L}^{\Lambda\rm CDM}   \left [1 + \sum_{i=1}^{N_\nu} \dfrac{\Delta_L^{(i)}}{2} \left(\tanh\left[\dfrac{\log(q_i)}{\Delta_q}\right]+1\right) \right],
\label{eq:fitbL_nus}
\ee
where at $z_{\rm coll}=0.7$, similarly to the light-relic case, we find that  $\Delta_L^{(i)} = 0.55\, f_{\nu_i}$, $q_i= 5 k/k_{\rm fs}^{(i)}$, where we have defined $f_{\nu_i} = \Omega_i/\Omega_m$, with $\Omega_i h^2 = m_{\nu_i}/(93.14\,\rm eV)$.
The slope of $\Delta_L \equiv \Delta b_L/b_L - 1 = 0.55\, f_{\nu}$ is consistent with that of Ref.~\cite{LoVerde:2016ahu}, found to be in agreement with simulations in Ref.~\cite{Chiang:2017vuk}.
Note that the scaling of the step in the bias, $\Delta_L$, is smaller than we found for light relics, where $\Delta_L = 0.6 \, f_X$, since here we are giving mass to an otherwise present (albeit massless) neutrino, as opposed to including a whole new particle in our analysis.
Additionally, we calculate the linear Eulerian bias with Eq.~\eqref{eq:b1_fit}, where now
\be
\mathcal T_m(k) = f_c \mathcal T_c(k) + \sum_{i=1}^{N_\nu}  f_i \mathcal T_i(k),
\ee
and we remind the reader that the subscript $c$ stands for CDM+b.

Adding massive neutrinos also causes a scale-dependent Eulerian bias, of size $f_\nu \equiv \sum_i f_{\nu_i}$, as we show in Fig.~\ref{fig:Supp_9eV}.
Part of this bias is caused by the inclusion of neutrinos in the matter budget, and part of it is due to the effect of neutrinos in the collapse of the haloes.
Both effects contribute with similar sizes, and
simply considering the bias with respect to cold dark matter does not result in a purely scale-independent bias, as we will explore later. 
As a consequence, we can read from Fig.~\ref{fig:Supp_9eV} that the $3\%$  suppression in the matter power spectrum caused by massive neutrinos is reduced to $1\%$ for the haloes, making the effect of neutrinos harder to observe in galaxy power spectra~\cite{LoVerde:2014pxa}.
In Ref.~\cite{Raccanelli:2017kht} it was shown that ignoring the scale-dependence of the bias is a safe approximation with current cosmological data, albeit it would induce biases with more-precise data from next-generation surveys.
With {\tt RelicFast} we can compute the linear bias quickly and precisely, so it would be possible to include a calculation of the bias in any cosmological search of neutrino masses.

Additionally, both matter and halo power spectra in Fig.~\ref{fig:Supp_9eV} show a bump at scales $k = 10^{-3}-10^{-2} \, h$ Mpc$^{-1}$, when including neutrinos. 
This result was expected for the matter power spectrum~\cite{Jimenez:2016ckl,Lesgourgues:2011rh}, but we see that the scale-dependence of the linear bias enhances the bump in the halo power spectrum to the half-percent level.
This enhancement is also present in the rest of the relics we have studied, although it is too small to warrant any further consideration.

\subsection{Neutrino hierarchies}

Let us now study the effect of the neutrino hierarchy on the halo bias and power spectrum.
For total neutrino masses $M_\nu \gg 0.1$ eV the precise difference between the neutrino mass eigenstates is largely irrelevant, and both hierarchies are well approximated as three degenerate neutrinos, with the same mass.
Let us, instead, study the opposite case, where $M_\nu=0.09$ eV. This is the lowest mass possible within the IH, where we will have two massive neutrinos, with $m^{\rm IH}_{\nu_i}=\{0.045,0.045\}$ eV. In the NH, however,  we will have three massive neutrinos, with masses $m^{\rm NH}_{\nu_i}=\{0.05,0.02, 0.02\}$ eV.
We compare the two hierarchies to the 3deg.~approximation, where $m_{\nu_i}^{\rm 3deg.} = \{0.3,0.3,0.3\}$ eV, and the 1$\nu$ case in Fig.~\ref{fig:Supp_9eV}.
This Figure shows that the 1$\nu$ approximation fails to reproduce either of the two hierarchies, as it overpredicts the amount of suppresion~\cite{Lesgourgues:2006nd}, and the more-massive single neutrino has a larger $k_{\rm fs}$, resulting in a displacement of the suppression to larger $k$.
However, taking three degenerate neutrinos, with the same $M_\nu =0.09$ eV, reproduces the bias for the NH to great precision at all scales, and only deviates from the IH within $0.1\%$ at intermediate scales, showing that the 3deg.~case is indeed a good approximation to both hierarchies.

\begin{figure}[h!]
	\includegraphics[width=0.44\textwidth]{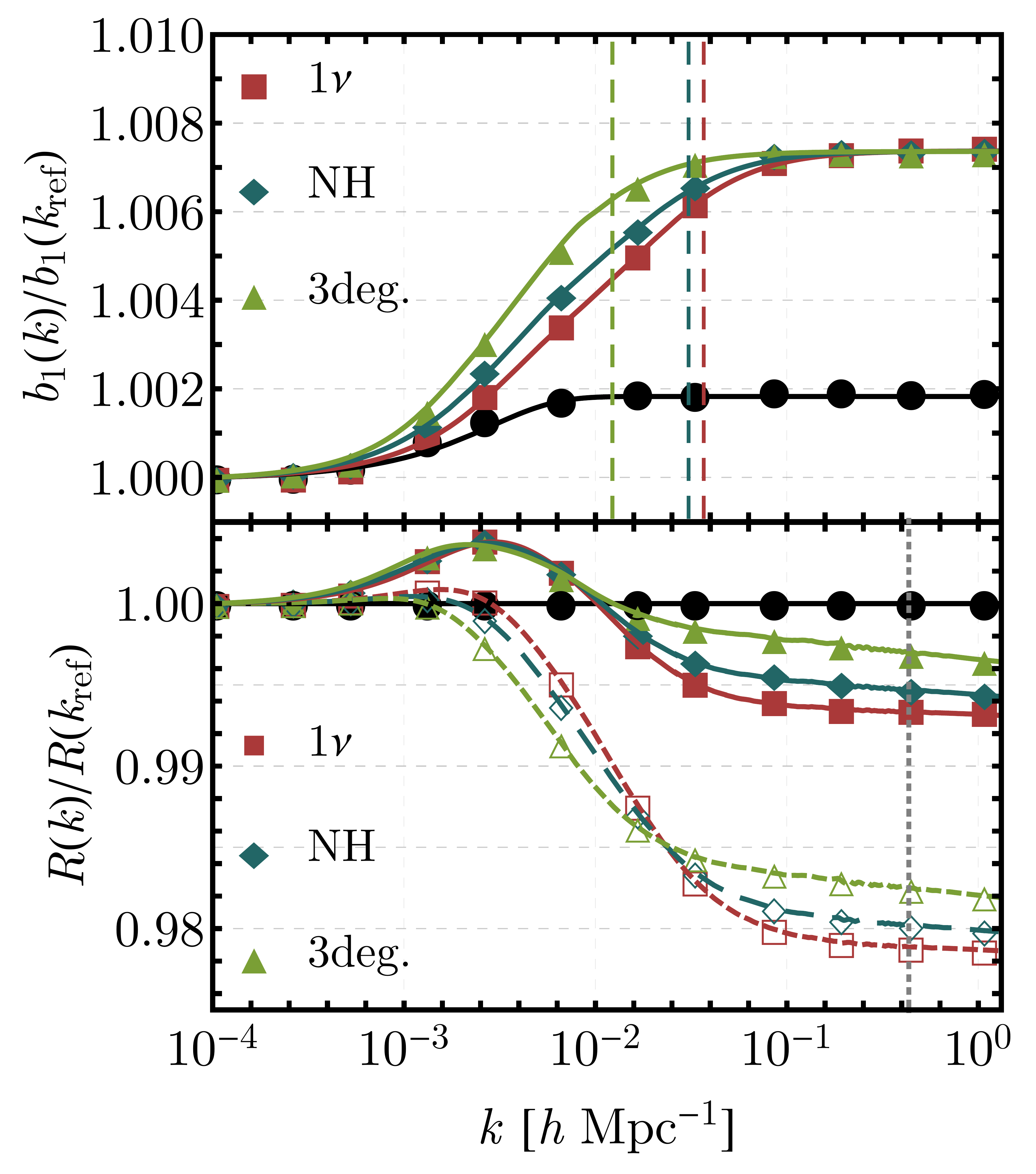}		
	\caption{Same as Fig.~\ref{fig:Supp_9eV}, albeit with $\sum m_{\nu_i} = 0.06$ eV.
	}
	\label{fig:Supp_6eV}
\end{figure}

We now study the case of $M_\nu=0.06$ eV, at the edge of the minimum neutrino mass possible (and thus only allowed by the NH). In this case the NH has two massive neutrinos, with $m_{\nu_i}=\{0.05,0.01\}$ eV, which we compare in Fig.~\ref{fig:Supp_6eV} with the 1$\nu$ approximation (with $m_\nu=0.06$ eV), and the 3deg.~case (with $m_{\nu_i}^{3\rm deg.}=\{0.02,0.02,0.02\}$ eV). 
We find that the suppression in the matter power spectrum is roughly 2\% for all cases, (with exact values of \{2.1, 2.0, 1.8\}\%  for 1$\nu$, NH, and 3deg., respectively, all at $k=1\,h$ Mpc$^{-1}$), whereas for the halo power spectrum this suppression is less pronounced, reaching values of  \{0.7,0.6,0.3\} \% for the same cases. Thus, the 1$\nu$ approximation is better at reproducing the NH, as expected, but the relative difference between these two cases is still of the order of 20\%.
This is to be expected, as these two cases have a distribution of neutrino masses that is also different by 20\%.
Additionally, we see that while for $M_\nu\gtrsim0.1$ eV the 3deg.~approximation works better than the 1$\nu$ case, this trend is reversed  at lower masses, so one should use the right hierarchy whenever possible.

Before moving on, let us correct some results from Ref.~\cite{LoVerde:2014pxa}.
Our formalism draws heavily from this reference, on which the scale-dependent bias induced by massive neutrinos was studied.
However, we find that the step in the Lagrangian bias is, to a good approximation, proportional to the total neutrino mass, regardless of how this mass is distributed amongst the neutrinos (see, for instance, Fig.~\ref{fig:bias_nus_cdm}).
This is in stark contrast to the result in Ref.~\cite{LoVerde:2014pxa}, where the 1$\nu$ and 3deg.~approximations yielded steps in the bias that differed by a factor of two.
As we advanced in Section~\ref{sec:Collapse}, the source of this discrepancy is the treatment of the neutrino-pressure fluctuations.
In Ref.~\cite{LoVerde:2014pxa} the sound speed of massive neutrinos was set to zero (and, therefore, pressure fluctuations were ignored).
Neutrinos are confirmed to have a nonvanishing sound speed~\cite{Audren:2014lsa,Ade:2015xua}, so we self-consistently compute the density and pressure fluctuations induced by a long-wavelength perturbation, by using Eq.~\eqref{eq:deltarho} and the adiabatic sound speed calculated with Eq.~\eqref{eq:cad}.
The effect of the sound speed is more significant for lighter neutrinos, for which $c_s^2\,(\approx w)$ can be important.
We have checked that under the same assumptions as Ref.~\cite{LoVerde:2014pxa}, namely, taking $c_{s,i}^2=0$, we recover their results, and the step in the Lagrangian bias due to light neutrinos (with $m_{\nu_i} \leq 0.1$ eV) is overestimated by a factor of two. 
We note, however, that in the separate-universe formalism
pressure fluctuations from all species are automatically included in the evolution of the long-wavelength CDM+b overdensity $\delta_L$, as shown in Ref.~\cite{Hu:2016ssz}. Therefore, one only needs to explicitly specify $c_s^2$ during the collapse when evolving $R(z)$, as we do, instead of $\delta_S(z)$. Nonetheless, the {\tt CLASS} code takes the same assumption of $c_s^2 = c_{\rm ad}^2$, so both formalisms should be equivalent.

We have also compared our results with those of Ref.~\cite{Chiang:2017vuk},  where the scale-dependent bias of massive neutrinos was computed both with N-body separate-universe simulations, as well as with different theoretical estimates.
As Ref.~\cite{Chiang:2017vuk}, we include $N_\nu=28$ neutrinos of $m_{\nu_i}=0.05$ eV, and fix the cosmological parameters to $h=0.7$, $\Omega_b h^2=0.0245$, $\Omega_d h^2=0.1225$, and a helium fraction of $Y_{\rm He}=0.24$. 
For practical purposes, instead of adding 28 massive neutrinos we make use of the equivalent-fermion approximation, outlined in Section~\ref{sec:Other_Relics}, and simply run {\tt RelicFast} with $\Omega_X h^2=0.015$ and $m_X=0.115$ eV, with no other neutrinos (massive or massless).
In this case we find steps in the bias of $\Delta b_1^L/b_1^L = 1.065$ and $\Delta b_1^L/b_1^L = 1.068$, at $z_{\rm coll}=0$ and 1, respectively, in excellent agreement with Ref.~\cite{Chiang:2017vuk}.
Note, however, that this step is not entirely due to the neutrino masses, since a large $N_\nu$ increases the amount of radiation in the early Universe, and thus the step in the Lagrangian bias, even if these neutrinos are massless. 
For instance, we estimate that setting $N_{\rm eff}=28$ (and thus considering 28 massless neutrinos) causes steps in the Lagrangian bias of $\Delta b_1^L/b_1^L = 1.01$ at $z=0$, and $\Delta b_1^L/b_1^L =1.02$ at $z=1$, which are a factor of $\sim4$ larger than the expected $\Lambda$CDM results for $N_{\rm eff}=3.046$.

\subsection{Bias with respect to Cold Dark Matter}

Throughout this work we have employed the usual definition of the linear Eulerian bias of Eq.~\eqref{eq:b1}, which includes the effect of light relics both in the spherical collapse and in the matter power spectrum.
This bias allows for a direct comparison of the halo power spectrum, as observed in galaxy surveys, and the matter power spectrum, inferred for instance through weak lensing.
One can choose to define the linear bias with respect to the cold dark matter, though, in which case it would be given by
\be
\tilde b_1 (k) \equiv \dfrac{P_{hc}(k) }{P_{cc}(k) } = \left[1+b_1^L(k) \right],
\ee
where the contribution of light relics as matter is removed, and the only scale dependence is through $b_1^L(k)$. We show this quantity, along with $b_1^L(k)$ for ease of comparison, in Fig.~\ref{fig:bias_nus_cdm}. Clearly, $\tilde b_1$ shows a smaller scale-dependent feature than $b_1$, as pointed out in Refs.~\cite{Villaescusa-Navarro:2013pva,Castorina:2013wga}, although the effect is still nonvanishing, as reported by Ref.~\cite{LoVerde:2014pxa}. 
From Fig.~\ref{fig:bias_nus_cdm} we find that, for $M_\nu=0.09$ eV, the amplitude of the $\tilde b_1$ step is $\Delta \tilde b_1/\tilde b_1 \approx 0.35\%$ for the haloes we consider, clearly tracing $b_1^L$, which shows a percent step due to neutrinos (half of which is already present in $\Lambda$CDM).
We have confirmed that for $M_\nu=0.06$ eV  this effect persists, with a 0.6\% step in $b_1^L$, and therefore $ \Delta \tilde b_1/b_1 \approx 0.3\%$ for the haloes we consider.

\section{Conclusions}
\label{sec:Conclusions}

In this paper we have presented the code {\tt RelicFast}, which can compute, through spherical collapse and the peak-background split, the bias of haloes in the presence of light relics, including neutrinos.
We have argued that this allows galaxy surveys to target light relics with masses above an meV, which comprise part of the matter today, and thus complements searches for relativistic relics, such as those of Refs.~\cite{Baumann:2017gkg,Baumann:2018qnt}, which target their effect on $N_{\rm eff}$.

Light relics can be any degree of freedom that decoupled from the standard model in the early Universe, and stayed relativistic. 
We have shown how all light relics with a Fermi-Dirac distribution can be expressed in terms of an equivalent neutrino with some mass and temperature. 
Using this insight, we have studied the effects in the halo power spectrum of an eV-scale sterile neutrino, with an arbitrary cosmic abundance, as suggested by short-baseline neutrino experiments~\cite{Dentler:2018sju}.
Additionally, we have shown that even bosonic degrees of freedom can be well approximated by a neutrino, if the neutrino temperature and mass are chosen wisely. We illustrated this by computing the matter and halo power spectra for both a scalar and a vector relic.
In all cases we have chosen relatively large values of the light-relic fraction $f_X$ to more clearly showcase their effect on the halo power spectrum, although given that both the suppression in power and the scale-dependent bias scale roughly linearly with $f_X$, our results can be easily translated to other relic abundances.

We have also explored the impact of massive neutrinos in the galaxy bias. We find that the linear bias is modified both by the inclusion of neutrinos in the cosmic matter budget, as well as by their effect on the spherical collapse of the haloes. Together, these effects cause a step-like linear bias of size $f_\nu$, which partially compensates the neutrino-induced suppression in the matter power spectrum.
In addition, we have shown how the effect on the galaxy power spectrum of both the normal and inverted hierarchies cannot be well represented by either three degenerate neutrinos, or a single one, for small neutrino masses ($M_\nu <0.1$ eV). It is, therefore, imperative to properly model each neutrino hierarchy for a robust detection---or constraint---of the neutrino masses.

Throughout this work we have only computed local biases to linear order. 
It would be interesting, however, to go beyond this approximation, in order to include larger $k$ modes with the necessary precision. This would require careful modeling of the neutrino and CDM fluids~\cite{Biagetti:2014pha,Perko:2016puo,Senatore:2017hyk,Castorina:2015bma},
and of their nonlinearities~\cite{Banerjee:2016zaa,Bird:2018all,Dakin:2017idt, AliHaimoud:2012vj}.
Additionally, galaxies are observed in redshift space, and the redshift-space distortions can be modified in the presence of light relics~\cite{Marulli:2011he,Kaiser:1987qv,Villaescusa-Navarro:2017mfx}.
Nonetheless, we do not expect the scale dependence of the linear bias to be significantly altered by any of these effects, and we leave their modeling for future work.

We have argued that
any search of neutrinos, or other massive relics using galaxy power spectra should include their effect on the galaxy bias, even if this effect is not observable at high significance in isolation.
Simulations have been used to obtain these biases~\cite{Bird:2018all,Villaescusa-Navarro:2017mfx}, although they are computationally prohibitive if cosmology or the relic properties are to be varied.
{\tt RelicFast} computes halo bias efficiently and accurately for any given cosmology including light relics, which allows for MCMC searches of these particles.
Additionally, constraints on the neutrino mass can be consistently achieved, by including all effects of neutrinos in the power spectra.
Therefore, we believe that {\tt RelicFast} holds great potential for the use of galaxy data.

\section*{Acknowledgements}

We wish to thank Emanuele Castorina, Marilena LoVerde, and Azadeh Moradinezhad for enlightening discussions and comments on a previous version of this manuscript, as well as Neal Dalal for useful correspondence, and Nina Maksimova for collaboration in the early stages of this project.
We are also thankful to the creators of both {\tt CLASS} and {\tt CAMB} for making their codes publicly available.
This work was supported by the Dean's Competitive Fund for Promising Scholarship at Harvard University.

\bibliography{bias_relics}{}
\bibliographystyle{bibpreferences}

\appendix

	\section{Use of the Code}
	\label{app:Code}
	
	In this Appendix we explain the main properties of the {\tt RelicFast} code, and how to best utilize it.
	{\tt RelicFast} is written in C++, and is parallelized using OpenMP.
	After a successful installation, {\tt RelicFast} will read a text file with the input parameters and output the Lagrangian and Eulerian biases to a text file.
	The inputs are the cosmological parameters (the relevant $\Lambda$CDM parameters, the neutrino masses, and the light-relic parameters),
	the $k$ values for which the bias is to be calculated,
	as well as the redshifts $z_{\rm coll}$ of collapse, and the masses $M$ of the haloes that are formed.
	
	The code runs {\tt CLASS} for the input cosmology (albeit it can be easily adapted for {\tt CAMB}). It then uses the transfer functions outputted to find the initial conditions for $R(z)$ (since $R_i$ and $\dot R_i$ depend on $\sigma(M)$), as well as for the evolution of the non-cold components inside the collapse equation.
	We evaluate the transfer functions at 100 values of redshift between 0 and $z_i=200$ (spaced as $\sqrt{z}$ for efficiency) and interpolate between them.
	As a note, we follow Ref.~\cite{LoVerde:2014pxa} in choosing $z_i=200$,  early enough that little nonlinear evolution has occured, but late enough that baryons have transferred most of the acoustic oscillations to the dark matter. 
	We have found that starting at different $z_i$ in the 100-400 range makes a relative change in $\Delta_L$ (defined as in Eq.~\eqref{eq:fitbL}) of a few percent, owing to both nonlinearities in the initial $\delta_S$ and the effect of baryon pressure~\cite{Naoz:2005pd}, which we do not include. 
	This should be treated as a lower bound on the uncertainty of our estimates.	
	
	We solve for the spherical collapse as explained in the main text, albeit using $z$ instead of time as the variable.
	The equation of motion is, thus,
	\ba
	&R''(z)  + R'(z)\left(\dfrac{1}{1+z} + \dfrac{H'(z)}{H(z)}\right)  = \\&- \dfrac{G [M + \delta M_X(z)]}{R^2(z) H^2(z)(1+z)^2} \nonumber \\ &- \dfrac{R(z)H_0^2}{2 H^2(z)(1+z)^2} \sum_i \tilde \Omega_i(z) [1+3 w_i + (1 + 3 c_{{\rm ad},i}^2)\delta_i ],\nonumber
	\end{align}
	where prime denotes derivative with respect to $z$, and
	where we have defined 
	\be
	\tilde \Omega_i(z) = 	\dfrac{\bar \rho_i(z)}{\rho_{\rm crit}}
	\ee
	with a tilde, to distinguish it from the $z=0$ values used throughout the text. 
	The $i$-th species long-wavelength fluctuation is calculated as
	\be
	\delta_i = \delta_L  \dfrac{\mathcal T_i(k,z)}{\mathcal T_c(k,z_{\rm ini})},
	\ee
	and its (adiabatic) sound speed as
	\be
	c_{{\rm ad},i}^2(z) = w_i(z) + \dfrac{w_i'(z) (1+z)}{3[1+w_i(z)]}.
	\ee

	Most of the numerical burden of solving this ODE is caused by the multiple interpolations of the transfer functions $\mathcal T_i$.
	We numerically solve this equation using Heun's method (a second-order Runge-Kutta method), as it improves accuracy dramatically over Euler's method, requiring no additional interpolations.
	We logarithmically bin in redshift to better sample low redshifts, where the halo evolves faster (note that this means that our $z=0$ results are actually at $z=10^{-2}$, to avoid changing to linear binning. We have confirmed that this small difference in redshift does not change any results).
	
\begin{figure}[hbtp!]
		\includegraphics[width=0.49\textwidth]{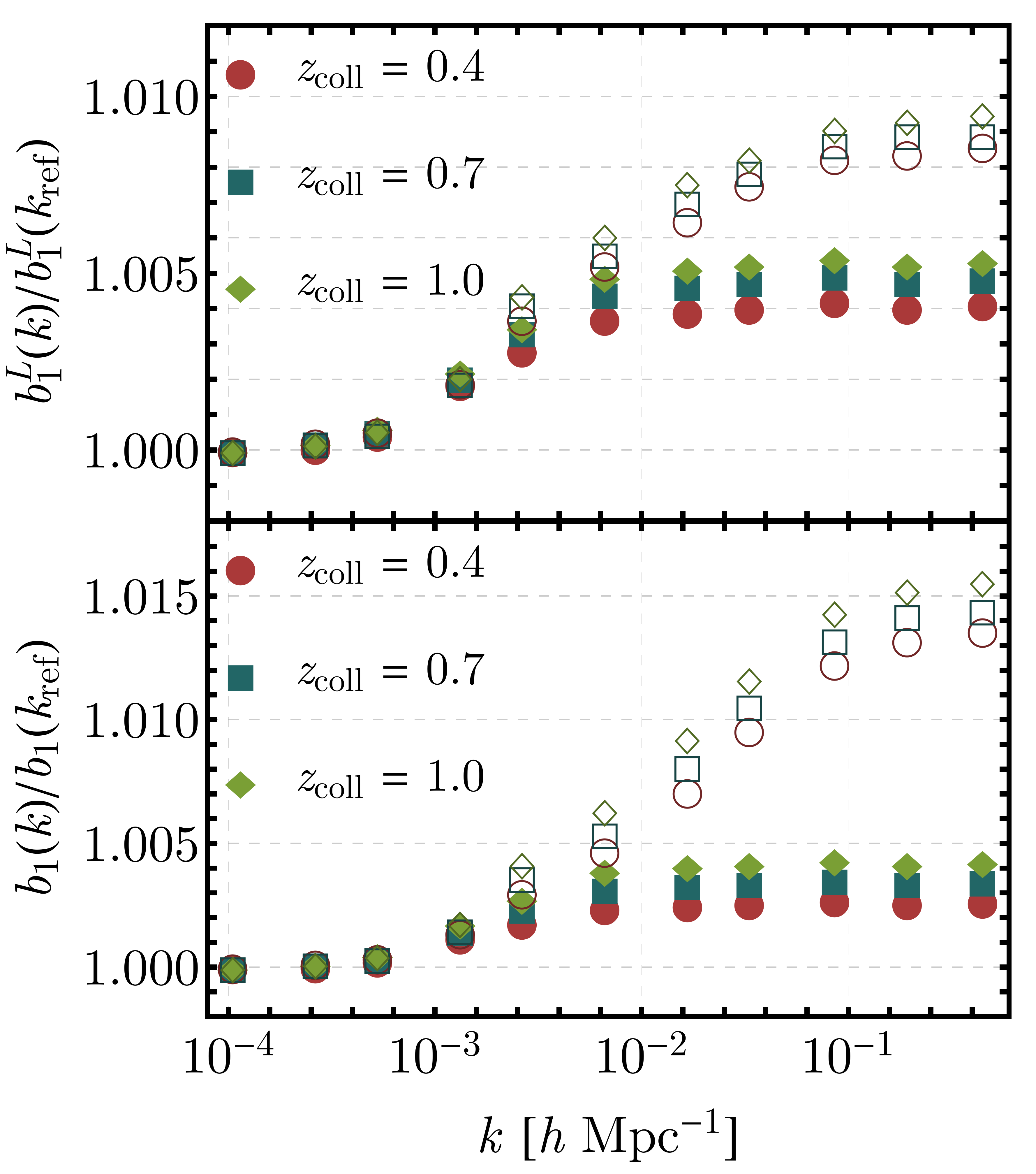}					
		\caption{Same as Fig.~\ref{fig:bias_Mhalo}, albeit fixing $M=10^{14}\,h^{-1}\,M_\odot$ and varying the redshift $z_{\rm coll}$ of collapse. The blue squares represent the same haloes here and in Fig.~\ref{fig:bias_Mhalo}.
		}
		\label{fig:bias_zcoll}
		\end{figure}

	
	The precision of {\tt RelicFast} can be manually altered by the user.
	We have found that the scale dependence of $b_1^L$ can be calculated at great accuracy even for moderate precision in its overall amplitude.
	This is because we are finding the change in $\delta_{\rm crit}$ when adding a long-wavelength perturbation $\delta_L$, and thus any overall rescaling of $\delta_{\rm crit}$ only modifies the absolute value of $b_1^L$, and not its scale dependence.
	Therefore, for best results, we encourage users to treat $b_1^L$ as the free parameter to marginalize over, instead of $b_1$. 
	Additionally, as shown in Fig.~\ref{fig:bias_Mhalo} for the case of a 0.1-eV neutrino, the scale dependence of $b_1^L$ is almost entirely mass-independent, as opposed to that of $b_1$, which depends moderately on redshift, and strongly on mass.
	We show the result of a similar analysis, albeit varying $z_{\rm coll}$ instead of $M$, in Fig.~\ref{fig:bias_zcoll}.
	This shows that the scale dependence varies comparably in $b_1$ and $b_1^L$ when changing the redshift of collapse.

	\section{Other Halo Mass Functions}
	\label{app:HMF}
	Here we discuss our bias results when using other HMFs, in particular those of Refs.~\cite{Bhattacharya:2010wy} (which was calibrated for $w$CDM cosmologies) and~\cite{Sheth:1999mn}. 
	In the first case, the HMF term is~\cite{LoVerde:2014pxa}
	\ba
	\dfrac{\partial \log n}{\partial \delta_{\rm crit}} =& \dfrac{q- (\sqrt{a(z)}\,\delta_{\rm crit}/\sigma)^2}{\delta_{\rm crit}} \\ \nonumber&- \dfrac{2 p}{\delta_{\rm crit}[1+(\sqrt{a(z)}\,\delta_{\rm crit}/\sigma)^{2p}]},
	\end{align}
	with $q=1.795$, $p=0.807$, and $a(z)=0.788\,(1+z)^{-0.01}$.
	Additionally, the more traditional ST HMF from Ref.~\cite{Sheth:1999mn} is well fit by the previous formula, albeit with $q=1$, $a=0.707$, and $p=0.3$.
	
\begin{figure}[b!]
	\includegraphics[width=0.49\textwidth]{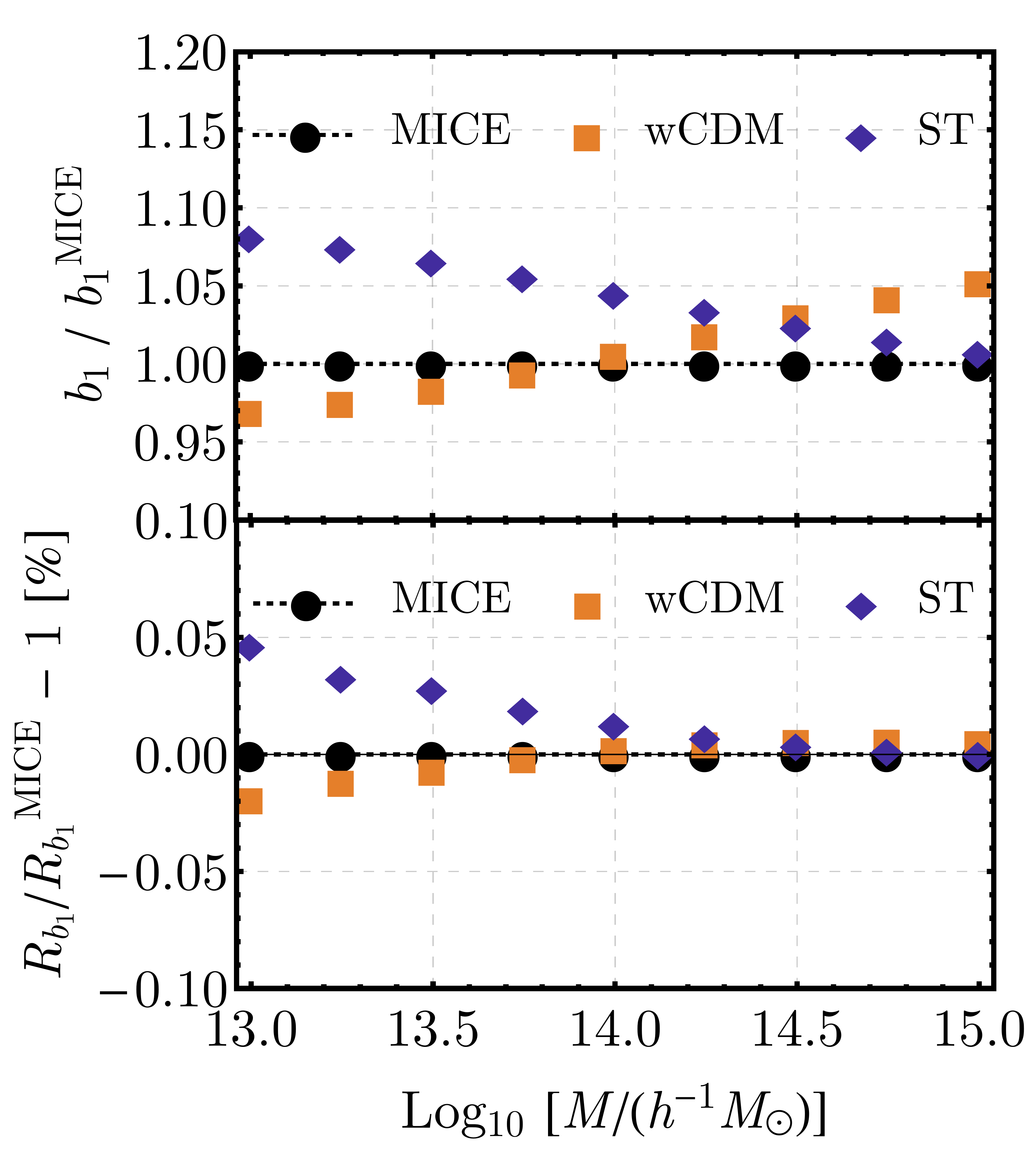}			
	\caption{We show the ratio of $b_1$ (the linear Eulerian bias) and $R_{b_1}$ (the step in $b_1$ caused by neutrinos) obtained with different mass functions, compared the one from MOCE (Eq.~\eqref{eq:HMF}), as a function of halo mass.
	In the bottom panel we have taken a cosmology with a 0.1-eV neutrino, and the step in the bias is computed at two different wavenumbers, $k_{\rm ref}=10^{-4}\,h$ Mpc$^{-1}$ and $k_s=1\,h$ Mpc$^{-1}$.
	}
	\label{fig:HMF_Eul_Rel}
\end{figure}

	In Fig.~\ref{fig:HMF_Eul_Rel} we show the the relative difference between the Eulerian biases when compared to our baseline case (MICE, from Ref.~\cite{Crocce:2009mg}), for haloes of different masses $M$, collapsing at $z=0.7$, all measured at $k_{\rm ref}=10^{-4}\,h$ Mpc$^{-1}$. The three HMFs agree well, and in particular the wCDM one (from Ref.~\cite{Bhattacharya:2010wy}) and MICE agree at the percent level.
	
	Nonetheless, the small discrepancies are very much scale independent. As we explained in the main text, the scale dependence of the Lagrangian bias is indifferent to the choice of HMF, which can only affect its normalization.
	A change in the overall value of $b_1^L$ can, however, modify the scale dependence of $b_1$.
	To find the size of this effect we define the quantity
	\be
	R_{b_1} = b_1(k_s) / b_1(k_{\rm ref}), 
	\ee
	as a measure of the step induced in the Eulerian bias by a light relic, where we have chosen the short-wavelength mode to be $k_s = 1\, h$ Mpc$^{-1}$, and the long wavelength mode at $k_{\rm ref} = 10^{-4}\,h$ Mpc$^{-1}$.
	To show that the choice of the HMF does not alter the step in the Eulerian bias at any appreciable level, we have run a case with a massive neutrino with $m_\nu=0.1$ eV (which yields a step of $R_{b_1}=1.01$, and similarly in the Lagrangian bias).
	We find that the percent-level differences in the overall values of the bias predicted by each HMF translate into 
	a difference in $R_{b_1}$ only at the 0.02-0.05\% level, decreasing for higher-mass haloes, as we show in Fig.~\ref{fig:HMF_Eul_Rel}.
	Thus, it is largely irrelevant which HMF to choose. In any case, this problem disappears if one marginalizes over the amplitude of $b_1^L$ instead of $b_1$, as the scale dependence is then independent of the chosen HMF.

\section{Clustering of Light Relics}
\label{app:Clustering}

\begin{figure}[h!]
	\includegraphics[width=0.44\textwidth]{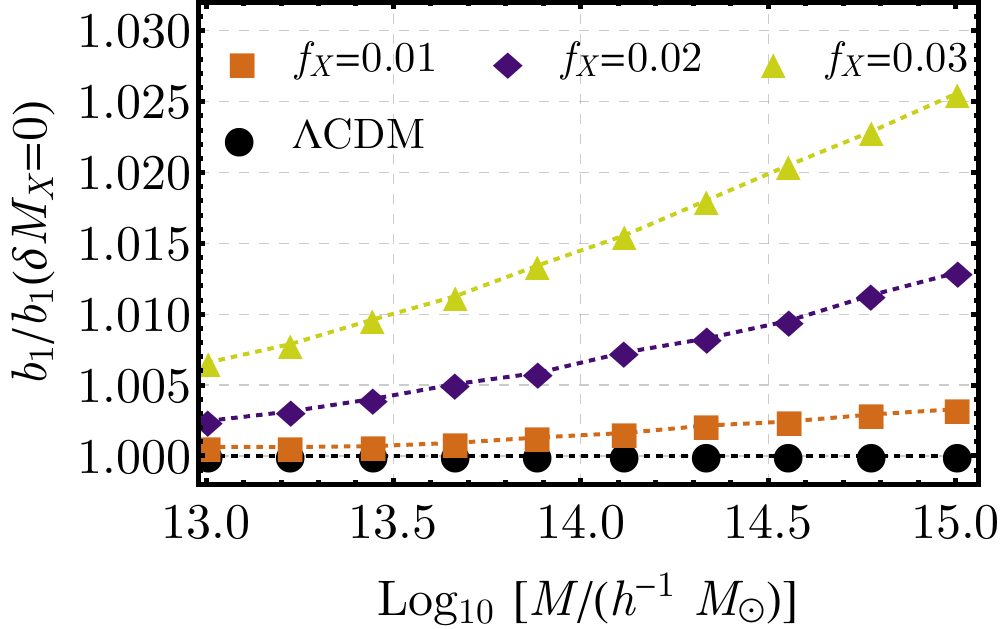}
	\includegraphics[width=0.44\textwidth]{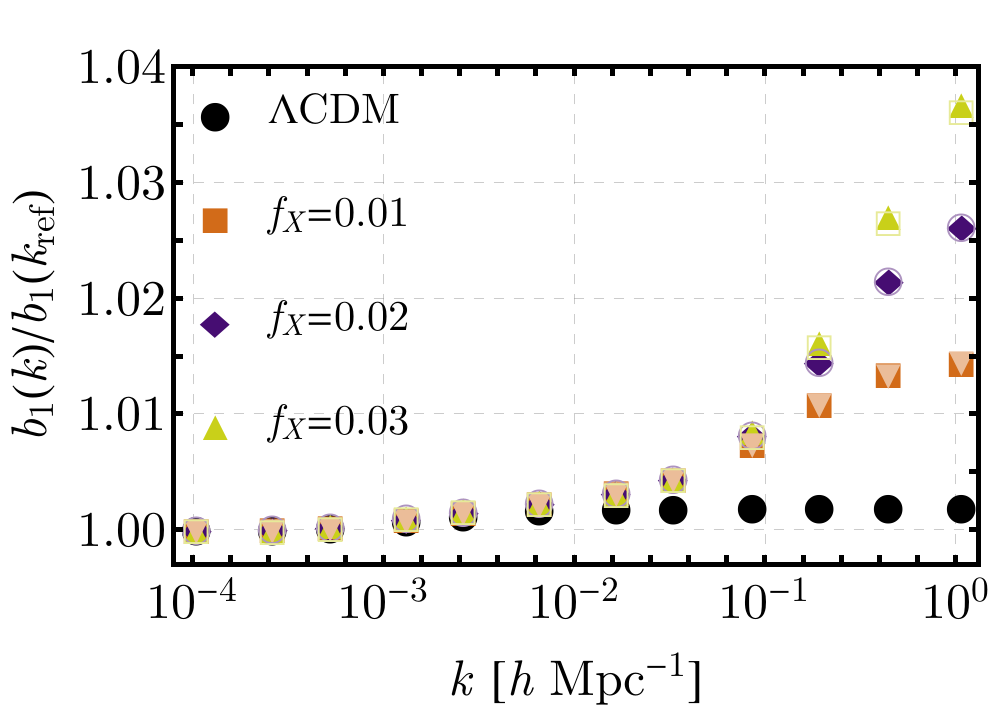}		
	\caption{
		{\bf Top:}	Shows the ratio of the linear bias with and without including collapse, for the same cases as Fig.~\ref{fig:bvsM_TR}.	
		{\bf Bottom:} Same as Fig.~\ref{fig:bvsk_TR_Neff_Lag}, where open symbols (and lighter colors) denote no-clustering.
	}
	\label{fig:Clustering}
\end{figure}

Here we detail the process of calculating the clustering of light relics, as well as how much it can affect our results. We emphasize that our focus is not on the properties of a light-relic halo surrounding the DM one, but instead we want to know its effect on the collapse time, and thus on the halo bias.	

In principle one should solve for the evolution of the DM and light-relic haloes collapsing simultaneously. However, the thermal speed of light relics for all cases we consider is rather large, so their clustering in the potential well of a halo can be expanded perturbatively. 
To estimate this clustering we will employ the Vlasov equation for a fermion~\cite{Ringwald:2004np,AliHaimoud:2012vj}, 
\be
\partial_\tau f + \dfrac{\mathbf p}{a m} \nabla_{\mathbf x} f - a m \nabla_{\mathbf x}\phi \cdot \nabla_{\mathbf p} f = 0,
\ee
where $\tau$ is the conformal time, $\phi$ is the gravitational potential perturbation, $\mathbf p = a(t) \mathbf q$ is the comoving momentum, and $\mathbf x = a^{-1} \mathbf r$ is the comoving position of a particle. 
We can now expand $f$ perturbatively in $\phi$, so
\be
f = f_0(p) + \delta f(\mathbf x,\mathbf p, t),
\ee
where $f_0$ is the unperturbed (Fermi-Dirac) distribution, and $\delta f$ is the first-order perturbation. We can thus use the``BKT" approximation of the equation for $\delta f$~\cite{Brandenberger:1987kf}
\be
\partial_\tau \delta f + \dfrac{\mathbf p}{a m} \cdot \nabla_{\mathbf x} \delta f - \dfrac{a m}{p} \nabla_{\mathbf x}\phi \cdot \mathbf p \dfrac{d f_0}{dp} = 0.
\ee
For the particular case of a top-hat mass overdensity, which is radially symmetric, this equation has an exact solution~\cite{LoVerde:2013lta}
\ba
\delta f(x, p, \mu,z) =& 2 m_X \dfrac{df_0/dp}{x^2} \int_{z}^{z_i} \dfrac{dz'}{H(z')} \delta M (z') \left( \dfrac{x_c}{x}-\mu \right) \nonumber \\
 &\times \begin{cases}
 	\dfrac{x^3}{x_{\rm halo}^3}, & \text{if $x_{\rm halo} > |\mathbf x - \mathbf x_c|$}.\\ \\
 	\dfrac{1}{ |\mathbf x - \mathbf x_c|^3}, & \text{otherwise},
 \end{cases}
\label{eq:intdf}
\end{align}
where $\mu=\hat x \cdot \hat p$ is the cosine of the angle between position and momentum, 
$\mathbf x_c \equiv [\eta(z)-\eta(z')] \mathbf p/m_X$ is related to the distance traveled by a particle, with $\eta(z)$ the superconformal time ($d\eta = dz (1+z)/H(z)$), and we explicitly compute
\be
|\mathbf x - \mathbf x_c| = \sqrt{x^2 + x_c^2 - 2 x x_c \mu}.
\ee
We have also defined  the comoving radius $x_{\rm halo}(z') = R_{\rm halo}(z') \, (1+z')$  of the halo,
and the (DM+b) mass overdensity $\delta M (z') = M_{\rm halo} - M_{\rm smooth}(z')$, with 
\be
M_{\rm smooth}(z') =\dfrac{4\pi}{3} R_{\rm halo}^3(z') \rho_{c}(z')  =  \dfrac{H_0^2  \Omega_{c} }{2} x_{\rm halo}^3(z').
\ee
Note that we require to know the solution for the halo collapse (i.e., $R(z)$) in order to calculate $\delta f$, which would require for us to solve both equations simultaneously. 
However, given that the change due to the clustering of light relics is a small perturbation, we can solve for $R(z)$ once without including it, and use that solution to find the clustering, which is then fed back to the collapse equation. 
We have attempted to perform this procedure iteratively, and found that after just one iteration it is converged better than one part in $10^4$.
We calculate the integral in Eq.~\eqref{eq:intdf} for a set of values of $x, p,$ and $\mu$ to be able to find the light-relic mass overdensity as
\be
\delta M_X (z) = \dfrac{m_X}{\pi} \int_0^{x_{\rm halo}(z)} \!\!\!\!\!\!\!\!\!\!\!\! dx x^2 \int_0^\infty \!\!\! dp p^2 \int_{-1}^1 d\mu \,  \delta f(x, p, \mu,z),
\ee
in natural units (note that if we had used $h$ instead of $\hbar$ we would have not had a $(2\pi)^{-3}$ factor in the $p$ integral).
This factor would enters the ODE we solve, Eq.~\eqref{eq:Rdotdot_Clust}, as explained before.

	\begin{figure}[h!]
	\includegraphics[width=0.45\textwidth]{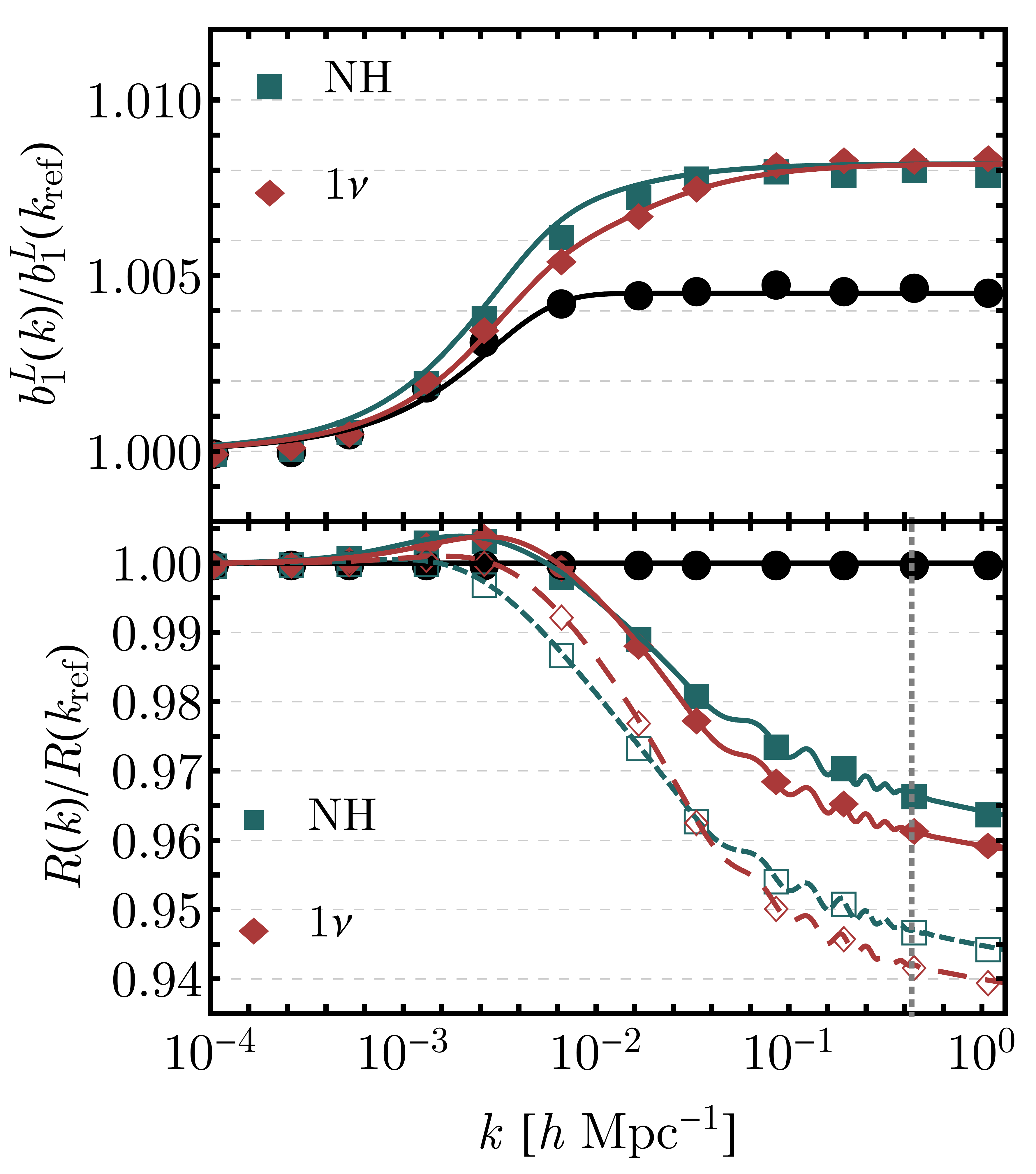}	
	\caption{
		Linear Lagrangian bias and suppression factors for the NH and the 1$\nu$ approximation, for the same parameters as in Fig.~\ref{fig:Supp_9eV} (i.e., $\sum m_{\nu_i}=0.09$ eV), although fixing $\Omega_m$ today instead of $\Omega_d$.
	}
	\label{fig:TR_nu_omegam}
\end{figure}

In Ref.~\cite{LoVerde:2014rxa} it was shown that the calculation here outlined, while producing a smaller neutrino halo, agreed with an N-1-body simulation on the collapse time, even for neutrinos as heavy as $m_\nu = 1$ eV, and haloes of $M=10^{15} \, M_\odot$. This comprises all the light-relic cases that we have studied in this paper for $M=10^{13} \, h^{-1} M_\odot$.
Moreover, we have seen that, while $\delta_{\rm crit}$ changes when considering light-relic collapse (see the top panel of Fig.~\ref{fig:Clustering}), the change is largely scale-independent, even in $b_1$, so whether we include clustering of light relics or not does not change our main results in this paper.
We show this in the bottom panel of Fig.~\ref{fig:Clustering}.
Therefore, for fastest results we suggest ignoring light-relic clustering.

\section{Fixing $\Omega_m$ versus $\Omega_d$}
\label{app:Omegam}

We show in Fig.~\ref{fig:TR_nu_omegam} the Eulerian bias and suppression factors for the case of neutrinos with $\sum m_{\nu_i}=0.09$ eV, both with the NH and with the 1$\nu$ approximation, where now we keep $\Omega_m$ fixed (by varying the DM density $\Omega_d$ when adding neutrinos).
The suppression in the matter power spectrum is more pronounced in this case, reaching a value of 6\%, as opposed to the 3\% we found in Fig.~\ref{fig:Supp_9eV}.
Similarly, in Fig.~\ref{fig:Supp_9eV} we found a 1\% suppression in the halo power spectrum, which grows to 4\% when varying $\Omega_d$, as read from Fig.~\ref{fig:TR_nu_omegam}.
The Lagrangian bias is, however, rather insensitive to whether $\Omega_d$ or $\Omega_m$ is kept fixed.
We find that for light relics the same result holds true: fixing $\Omega_d$ results in a smaller suppression in the matter power spectrum, but does not significantly change the biases we calculate. Throughout this work we have kept $\Omega_m$ fixed for non-neutrino light relics, to simulate a CMB prior, although one is free to change either.

Additionally, in the case of neutrinos, changing $\Omega_d$ to keep $\Omega_m$ fixed results in wiggles in the curves of Fig.~\ref{fig:TR_nu_omegam}.
This is because of the change in the sound horizon, as we explained in Section~\ref{sec:Relics_bias} for the case of light relics.
In this case no new degrees of freedom were added to the cosmological model, so the phase of the BAO is unchanged, which is why no wiggles were found in Figs.~\ref{fig:Supp_9eV} and \ref{fig:Supp_6eV}, where we kept $\Omega_d$ fixed.

\end{document}